\documentstyle[12pt,epsf,amstex]{article}
\def\fun#1#2{\lower3.6pt\vbox{\baselineskip0pt\lineskip.9pt
\ialign{$\mathsurround=0pt#1\hfil##\hfil$\crcr#2\crcr\sim\crcr}}}

\newcommand{\be}{\begin{eqnarray}}
\newcommand{\ee}{\end{eqnarray}}
\newcommand{\bd}{\begin{displaymath}}
\newcommand{\ed}{\end{displaymath}}
\newcommand{\ba}{\begin{array}}
\newcommand{\ea}{\end{array}}
\newcommand{\bt}{\begin{tabular}}
\newcommand{\et}{\end{tabular}}

\renewcommand{\theequation}{\thesection.\arabic{equation}}

\input epsf
\newcommand{\grpicture}[1]
{
    \begin{center}
        \epsfxsize=350pt
        \epsfysize=0pt
        \vspace{-5mm}
        \parbox{\epsfxsize}{\epsffile{#1.eps}}
        \vspace{5mm}
    \end{center}
}
\fontsize{12}{25}
\begin{document}

\vspace{.5cm}

\begin{center}
{\large
 Normalized vacuum  states in ${\cal N}  = 4$  supersymmetric Yang--Mills 
quantum mechanics with any gauge  group.}

\vspace{.5cm}

 {\bf V. G. Kac} \\
{\it IH\'ES, Le Bois--Marie, 35 route de Chartres, F-91440,
Bures-sur-Yvette, France \\
and \\
Departement of Mathematics, M.I.T., Cambridge, Massachusetts 02139, USA }\\ 
\ \ \\
and  {\bf A.V. Smilga}\\
{\it Universit\'e de Nantes, 2, Rue de la Houssini\`ere, BP 92208, F-44322,\\
Nantes CEDEX 3, France \\
and\\
ITEP, B. Cheremushkinskaya 25, Moscow 117218, Russia}

\end{center}

\vspace{.5cm}

\begin{abstract}
 
We study the question of existence and the number of normalized vacuum states
in ${\cal N} =4$ super--Yang--Mills quantum mechanics for any gauge group. 
The mass
deformation method is the simplest and clearest one. It allowed us to 
calculate the number of normalized vacuum states for all  gauge
groups. For all unitary groups, $\#_{\rm vac} = 1$, but for the symplectic
groups [starting from $Sp(6)$ ], for the orthogonal groups
[starting from $SO(8)$]
and for all the exceptional groups, it is greater than one. We also discuss at 
length
the functional integral method. We calculate the ``deficit term'' for some 
non--unitary groups and predict the value of the integral giving the
``principal contribution''. The issues like the Born--Oppenheimer procedure
to derive the effective theory and the manifestation of the localized
vacua in the asymptotic effective wave functions are also discussed.

\end{abstract}

\section{Introduction}

Consider the theory obtained by the dimensional reduction of ${\cal N} = 4$,
$D=4$ super--Yang--Mills theory (which is obtained in turn by the
dimensional reduction of  ${\cal N} = 1$,
$D=10$ SYM theory to four dimensions) when the whole space is shrinked to
a point and we are dealing with a supersymmetric quantum mechanical (SQM)
 system 
involving 16 real supercharges. In (9+1)--dimensional notations,
\be
\label{Q}
Q_\alpha \ =\ \frac 1{\sqrt{2}} \left[ (\Gamma_I)_{\alpha\gamma} E_I^A
+ \frac g2 (\Gamma_I \Gamma_J)_{\alpha\gamma} f^{ABC} A_I^B A_J^C \right] \lambda^A_\gamma
\ee
\be
\label{H} 
H = \frac 12 E_I^A E_I^A + \frac {g^2}4 
f^{ABE} f^{CDE} A^A_I A^B_J A^C_I A^D_J +
\frac{ig}2 f^{ABC} \lambda^A_\alpha (\Gamma_I)_{\alpha\beta} \lambda_\beta^B
 A^C_I \ ,
\ee
where $E_I^A = -i \partial/\partial A_I^A$, 
$A = 1,\ldots,{\rm dim}(G)$, $I,J = 1, \ldots, 9$; $f^{ABC} $ are structure 
constants 
; $\lambda^A_\alpha$
are Majorana spinors forming the {\bf 16}--plet of $SO(9)$, 
they should be understood as quantum fermion operators satisfying 
$\{\lambda_\alpha^A, \lambda_\beta^B\}_+ = \delta^{AB} \delta_{\alpha\beta}$;
 $\Gamma_I$ are 9--dimensional (real and symmetric) $\Gamma$--matrices, 
$\{\Gamma_I, \Gamma_J\}_+ = 2\delta_{IJ}$.

The operators (\ref{Q}) and (\ref{H}) act on the wave functions depending on
$A_I^A$ and holomorphic fermion variables $\mu_{\tilde \alpha}^A,\ \ 
\tilde \alpha = 1,\ldots, 8$. We may choose
  \be
 \label{mulam}
\left\{
\begin{array}{c}
\mu_1^A = (\lambda_1^A + i\lambda_9^A)/\sqrt{2} \\ \ldots \\ \mu_8^A = 
(\lambda_8^A +
i\lambda_{16}^A)/\sqrt{2} \end{array} \right.
\ \  {\rm and} \ \ \left\{ \begin{array}{l}
\lambda_1^A \ =\  (\mu_1^A + \bar \mu_1^A)/\sqrt{2} \ =\ 
 \left( \mu_1^A + \frac {\partial}{ \partial  \mu_1^A }\right)
/\sqrt{2} \\
\ldots \\ 
\lambda_{16}^A \ =\ - i (\mu_8^A - \bar \mu_8^A)/\sqrt{2} \ =\ 
 - i\left( \mu_8^A - \frac{ \partial}{ \partial  \mu_8^A} \right)/\sqrt{2}
\end{array} \right.
\ .
 \ee
The specifics of the ${\cal N} =4$ theory (compared with ${\cal N} =1$
and ${\cal N} = 2$ theories) is that the hamiltonian (\ref{H}) does not 
conserve the fermion charge which is related to the fact that our holomorphic
variables $\mu_{\tilde \alpha}$ do not form a representation of $SO(9)$.

The dynamic variables $A_I^A\ , E_I^A\ , \lambda_\alpha^A$ 
are dimensionless (if relating as we will do later the SQM model 
(\ref{Q},\ \ref{H}) to a field
theory placed in a small box, $A_I^A$ are measured in the units of its inverse
size $L^{-1}$ and $\lambda_\alpha^A$ are measured in the units of
$L^{-3/2}$), but we have chosen not to rescale away the coupling constant
$g$. It sets up a characteristic energy scale $E_{\rm char} \sim g^{2/3}$.

The relation
 \be
\label{QQHG}
\{Q_\alpha, Q_\beta\}_+   \ =\ \delta_{\alpha \beta} H +  \frac g2
(\Gamma_I)_{\alpha\beta} A_I^AG^A 
 \ee
holds where 
  \be
 \label{Gauss4}
G^A = f^{ABC}\left(A_I^BE_I^C  - \frac i2 \lambda_\alpha^B \lambda_\alpha^C
\right)
  \ee
is the Gauss law constraint. We are interested only
in the gauge invariant states $G^A |\Psi \rangle = 0$ for which the second term
in (\ref{QQHG}) vanishes, and we have the standard algebra of extended SQM.
The dynamics of this theory (and more simple SQM theories with 8 and
4 real supercharges obtained by 
dimensional
reduction of ${\cal N}=2$ and ${\cal N}=1$ SYM theories) was a subject of intense interest 
since the middle
of eighties \cite{Halp,jaIW}.  As was first noted in \cite{jaIW}, the spectrum
of the hamiltonian (\ref{H}) is continuous and the band of delocalized states
starts right from zero. The reason for that is very simple. The classical 
potential energy in the hamiltonian (\ref{H}) goes to zero if 
\be 
\label{AIJ}
 f^{ABC}A_I^A A_J^B \ =\ 0
\ee
for all $I,J$.
The condition (\ref{AIJ}) means that $A^A_I t^A$ belong to the Cartan
subalgebra. Up to a global gauge transformation,  
 \be
\label{valley}
(A_I^A)_{\rm class.\ vac.} \ =\ A_I^s\ , \ \ \ \ \ \ 
  \ee  
 $s = 1,\ldots, r$,
where $r$ is the rank of the gauge group.
Back in 1982 Witten  noticed that, in supersymmetric case, 
this valley is 
not lifted by quantum corrections \cite{Wit}. As a result, the low energy 
wave functions
tend to smear out along the valley. As the valley  (or alias, the 
vacuum moduli space) 
(\ref{valley}) is not compact, the motion is infinite, the wave function
is delocalized, and the spectrum is continuous.

One can make this statement more accurate, writing down the supercharges and 
hamiltonian describing the motion along the valley of slow variables in 
Born--Oppenheimer spirit. In the lowest order in the Born--Oppenheimer
expansion parameter $1/(g|{\bf A}|^3)$, the result is very simple 
\cite{Wit,Trieste}
  \be
\label{QHeff}
Q_\alpha^{\rm eff} &=& \sum_{s=1}^r\frac 1{\sqrt{2}} (\Gamma_I)_{\alpha\beta} 
\lambda_\beta^s E_I^s \nonumber \\
H^{\rm eff} &=& \sum_{s=1}^r \frac 12 E_I^s E_I^s\ \ ,\
 \ee
if the orthonormal basis in the Cartan subalgebra is chosen (for clarity, the
sum over $s$ is written explicitly).
Thereby, the problem is reduced to the problem of free motion in the 
$(D-1)r = (2{\cal N}+1)r$--dimensional flat space [with a certain discrete 
symmetry imposed; this symmetry will be discussed in details in
the  Appendix, and  
a detailed derivation of Eq.(\ref{QHeff}) will be given in Sect. 4 ]. 
The spectrum is obviously continuous.  

The theory (\ref{H}) is interesting by itself, but also because of its 
relations to brane dynamics. The hamiltonian (\ref{H}) for the gauge group
$SU(n)$ in the large $n$ limit just coincides
with the mass operator of 2+1 supermembranes embedded in 9+1 - dimensional
space  \cite{Hoppe}. The fact that the spectrum of (\ref{H}) is continuous
means that the supermembrane mass spectrum is continuous \cite{Trieste,Nic}.
 The realization of this fact has quenched  early  attempts to
build up a supermembrane theory (where supermembranes were treated
 as fundamental objects ). 

The revival of  interest to the hamiltonian (\ref{H}) was due to
a recent discovery that on top of delocalized continuum spectrum states, the
hamiltonian (\ref{H}) enjoys also a {\it normalized} vacuum state.
\footnote{
The effective hamiltonian in Eq.(\ref{QHeff}) does not enjoy such a normalized 
vacuum.
There is no contradiction here because
 the supersymmetric vacuum state of the full hamiltonian 
(\ref{H}) is localized in 
the region $g|{\bf A}|^3 \sim 1$ which is just the region where 
the effective theory (\ref{QHeff}) makes no sense. An important remark, 
however, is that the existence of the normalized vacuum in the full theory
(\ref{H}) {\it can} be conjectured by analyzing the dynamics of the effective
free theory \cite{Halp1,Hoppe1}. We will discuss it in details in
Sect. 5 of the paper. } 
The 
existence of such a state is very important for $D$--brane theory (in the
modern approach where  $D$--branes are not believed to be fundamental 
ingredients of the theory, but kind of solitons in the holy grail M--theory).
In the following, we will not use the brane terminology, however, and will
concentrate on studying the dynamics of the hamiltonian (\ref{H}) as it is.

Originally, the existence of the normalized supersymmetric vacuum state
was demonstrated when calculating carefully the Witten index
 \be
\label{IWdef}
I_W \ =\ \lim_{\beta \to \infty} {\rm Tr} \left\{ (-1)^F e^{- \beta H} \right\}
\ =\ n_B^0 - n_F^0
 \ee
for the hamiltonian H. For the systems where the spectrum is discrete and where
all the states are localized, this is a rather straightforward method. As all
the bosonic states with non-zero energy have their fermionic counterparts, the
expression $ {\rm Tr} \{ (-1)^F e^{- \beta H} \}$ does not
depend on $\beta$ in this case. One can present ${\rm Tr} \{ (-1)^F
 e^{- \beta H} \}$ in the functional integral form and calculate it in the 
limit $\beta \to 0$ where the functional integral is reduced to a 
finite--dimensional integral of $ \exp \{- \beta H_{\rm cl}\}$ over the 
classical
phase space \cite{Cecotti}:
  \be
 \label{CG}
  I_W \ =\ \int \prod_n \frac {dx_n dp_n}{2\pi} \prod_a d\bar\psi_a d\psi_a
e^{-\beta H_{\rm cl}(x_n,p_n;\ \bar\psi_a,\psi_a)} 
 \ee
 For example, for the supersymmetric oscillator
\be
 \label{Hosc}
H \ =\ \frac{p^2}2 + \frac{\omega^2 x^2}2 + \omega \bar \psi \psi\ ,
 \ee
and
 \be
\label{intind}
I_W \ =\ \int \frac {dx dp}{2\pi} d\bar\psi d\psi
e^{-\beta H_{\rm cl}(x,p;\bar\psi,\psi)} \ = \ 1
 \ee
signalizing the presence of one bosonic vacuum state annihilated by the action 
of the supercharges and hamiltonian.

For systems with continuous spectrum, the situation is more intricate. Instead
of the discrete sum
 \be
 \label{superTr}
 {\rm Tr} \left\{ (-1)^F e^{- \beta H} \right\} \ =\ 
\sum_n (-1)^{F_n} e^{-\beta E_n} 
 \ee 
we have continuous integrals involving boson and fermion spectral densities
which, generally speaking, are not equal even though the hamiltonian
is supersymmetric, and we cannot argue anymore that the supertrace
${\rm Tr} \{ (-1)^F e^{- \beta H} \}$ is $\beta$--independent. We can write
 \be
\label{princdef}
I_W \ =\ {\rm Tr}_{\beta = \infty} \ =\  {\rm Tr}_{\beta = 0} \ +\ 
\int^\infty_0 d\beta\ \frac \partial{\partial \beta} {\rm Tr}_{\beta} 
\equiv I_W^p - I_W^d
 \ee
The first term here is called the ``principal contribution'' and the second
term is known as the ``deficit term''. In many cases, the second term
is just zero even though the spectrum is continuous. The example of
such ``benign'' system is the spectrum of massless Dirac operator 
on $R^4$ in the instanton background
(as it is well known, the chiral symmetry of this problem can be presented
as supersymmetry \cite{Gaume}).
As the instanton field falls away rapidly at large distances, we have the 
continuum spectrum states with asymptotics of plane waves. But on top of 
that, we also have the normalized fermion zero modes. Their number
is given by the Atiyah--Singer theorem, and the Atiyah--Singer index is
nothing else as the Witten index in this particular problem. It can
be presented in the form (\ref{princdef}) where the deficit term is absent.
The benign nature of the Dirac system is related to the fact that we can
compactify $R^4$ on $S^4$ (so that the spectrum becomes discrete)
while preserving the supersymmetry.

But it {\it is} not so for the problem under consideration. Both principal and 
deficit term contribute on equal footing here. Let us briefly comment
first on the calculation of the principal term. To begin with, assume that
the gauge group is $SU(2)$ (the most simple case). One can show that the 
corresponding finite dimensional integrals (for ${\cal N}=1$, ${\cal N}=2$, and
${\cal N}=4$ theories) have the form 
   \be
 \label{indprinc}
  I_W = \frac{1}{8\pi^2}  \left(\frac{\beta g^2}{2\pi} 
\right)^{3(2{\cal N}+1)/2}
\int \prod_{A\mu} dA^A_\mu \ \det \| i A^A_\mu \Gamma_\mu \epsilon^{ABC}\|
\nonumber \\
\exp\left\{ -\frac {\beta g^2}4 
\epsilon^{ABE} \epsilon^{CDE} A^A_\mu A^B_\nu A^C_\mu A^D_\nu \right \}\ 
= \ \ \frac{2^{2({\cal N}-2)} \Gamma({\cal N} - 1/2)}
{\sqrt{\pi} \Gamma({\cal N} + 1)}
 \ee
with $\mu = 0,1,2, 3$,  $\mu = 0,1,\ldots, 5$, and  $\mu = 0,1,\ldots, 9$
for ${\cal N}=1,\ {\cal N}=2$ and ${\cal N}=4$, respectively.
The integrals in the R.H.S. of Eq.(\ref{indprinc}) were first correctly
calculated in Ref.\cite{jaIW}. However, the correct expressions 
(\ref{indprinc}) differ from the expressions quoted in Ref.\cite{jaIW}
by the overall factor $1/4$ in the case of ${\cal N}=1$ theory, and by
$1/8$ in the case of ${\cal N}=2$ and ${\cal N}=4$ theories. Correspondingly, the correct
results (found in \cite{YiSethi}):
  \be
\label{princres2}
\left( I_W^{\rm p } \right)_{{\cal N}=1} = \frac 14,\ \ 
\left( I_W^{\rm p } \right)_{{\cal N}=2} = \frac 14,\ \ 
\left( I_W^{\rm p } \right)_{{\cal N}=4} = \frac 54
  \ee
differ from the results quoted in Refs.\cite{jaIW} by these factors.
\footnote{This is not just an arithmetic error. The difference in normalization
factor stems from different methods used when deriving (\ref{indprinc}). 
The authors of Ref.\cite{YiSethi} implemented gauge invariance by imposing
Gauss law constraint on the wave functions while the starting point
in Ref.\cite{jaIW} was the hamiltonian with the constraints explicitly
resolved on the classical level. The missing factors
can be restored in this approach if taking into account  the condition
of discrete Weyl invariance for the wave functions. See Appendix
for more details.}

The calculation of the principal contribution for more complicated groups
is a rather intricate business. For higher unitary groups $SU(n)$, it was 
done in recent \cite{Nekr}. The result is
   \be
   \label{princresn}
\left( I_W^{\rm p } \right)_{{\cal N}=1} =  
\left( I_W^{\rm p } \right)_{{\cal N}=2} = \frac 1{n^2}, \ \  
\left( I_W^{\rm p } \right)_{{\cal N}=4} = 1 +  \sum_{m|n} \frac 1{m^2}
 \ee
where  the sum in the last formula runs over all integer divisors $m$ of $n$ 
including $n$, but not including 1.

The results (\ref{princres2}, \ref{princresn}) are fractional, but the
number of supersymmetric normalizable vacuum states is, of course, integer.
That means that in our case the ``deficit term'' cannot be zero. And it is
not. A not so difficult calculation (which we will dwell upon in details
  later) displays
 \be
\label{defres}
(I_W^{\rm d})_{{\cal N} = 1} \ =\ (I_W^{\rm d})_{{\cal N} = 2}\ =\ 
\frac 1{n^2}\ , \nonumber \\
(I_W^{\rm d})_{{\cal N} = 4} \ =\ \sum_{m|n} \frac 1{m^2}
 \ee
for the $SU(n)$ gauge group. 
Subtracting (\ref{defres}) from (\ref{princresn}), we finally obtain
 \be
\label{sunres}
\left( I_W \right)_{{\cal N}=1} =  
\left( I_W \right)_{{\cal N}=2} = 0, \ \   
\left( I_W \right)_{{\cal N}=4} = 1 \ ,
 \ee
i.e. the quantum mechanical systems obtained by the dimensional reduction
of ${\cal N}=1$ and ${\cal N}=2$ supersymmetric Yang--Mills theories with $SU(n)$ gauge
group does not have a normalizable supersymmetric vacuum state at all,
while the theory (\ref{H}) has exactly {\it one} such state.

This is fine, but the method just outlined has two disadvantages. First, it
is indirect and does not give a clue how the wave function of the normalized
vacuum state looks like. Second, it is far from being obvious how to generalize
this method to non-unitary groups. The calculation of the primary contribution
is especially tricky. Already the paper \cite{Nekr} where the principal
contribution was calculated for unitary groups was technically very difficult.
We have no idea how to generalize it to symplectic, orthogonal, or exceptional
groups.
 What we {\it are} able to do is to calculate the deficit term
for other groups. We will present these calculation in Sect. 4.

Our main message, however, is different. As has been observed in 
\cite{Porrati}, the presence of the normalized vacuum state in ${\cal N}=4$ theory
 can be established without coming to grips with difficult calculations
of the integrals for Witten index. It suffices to deform the theory
adding the mass term to the matter fields (we are thinking now of our
theory in ${\cal N}=1$ 4-dimensional terms where it involves the gauge multiplet
and 3 chiral matter multiplets in the adjoint representation of the group).
 Establishing the  supersymmetric vacua becomes now an almost trivial 
business of solving some simple
algebraic equations.  The number of quantum vacua just coincides
with the number of (gauge--inequivalent) solutions of these classical
equations. It turns out that for unitary groups, there is only
{\it one} such solution. If the mass is large, the Born--Oppenheimer
approximation works, and one can just write down the vacuum wave function
explicitly. We are interested in the theory in the opposite limit
$M \to 0$ where this cannot be done. But, by continuity, the normalizable
vacuum state exists for any mass, however small it is. It is a natural
{\it hypothesis} that the state remains to be normalizable also at the point
$M = 0$. This  hypothesis is confirmed by the indirect calculations
of $I_W$. 

The great advantage of this mass deformation  method is that
its generalization  to higher groups is not difficult.
It turns out that the problem is reduced to the problem of classification
of the so called ``distinguished'' nilpotent elements of a complex simple Lie
algebra the solution of which has been
known to mathematicians for a long time. For higher simplectic [starting
from $Sp(6)$], higher orthogonal [starting from $SO(8)$] and for all 
exceptional
groups, the solution is not unique, and there are several supersymmetric
normalized vacuum states. 

In the next section, we will write down the equations determining
the positions of the classical vacua in the large $M$ limit, solve
them for the unitary groups, and discuss at some length the philosophy of
this method.
Sect. 3 is  devoted to solving these equations for other groups.
The final result for the counting of supersymmetric vacuum states
presents the content of Theorem 6.
In Sect. 4, we calculate  the deficit term in 
Eq.(\ref{princdef}) for some groups.  
In Sect. 5, we discuss a third way to deduce the existence of the normalized
vacuum state(s) in ${\cal N} = 4$ SYM quantum mechanics: via studying
the asymptotic solutions for the effective theory. We present a {\it simple}
derivation for the asymptotic supersymmetric wave function in the $SU(2)$ 
case.
Calculation of the principal contribution to the Witten index in the $SU(2)$
theory is the subject of the Appendix.

\section{Mass deformation of the ${\cal N}=4$ theory.}
\setcounter{equation}0

As we have mentioned before, the problem (\ref{Q}, \ref{H}) has exciting
reverberations for $D$--branes and M--theory. However, the dynamics
of this quantum mechanical model can be understood better if exploiting
the other relation: the relation of (\ref{Q}, \ref{H}) and of the conventional
4--dimensional supersymmetric field theory. In 4--dimensional language,
the variables $A^A_{1,2,3}$ present the zero Fourier harmonics of the gauge
fields, while the variables $A^A_{4,\ldots,9}$ are associated with matter
fields.  
Let us define
 \be
 \label{scalar}
\phi_1^A = A_4^A + i A_5^A,\ \ \phi_2^A = A_6^A + i A_7^A,\ \ \phi_3^A = A_8^A
 + i A_9^A
 \ee
These fields together with 12 (out of 16) components of of $\lambda_\alpha^A$
form 3 chiral matter ${\cal N}=1$ multiplets $\Phi_f^A$
($f = 1,2,3$ is the ``flavor'' index). In the matter sector, the lagrangian of
the ${\cal N}=4$ SYM theory presents the Wess--Zumino model with superpotential 
$\sim \epsilon_{fgh} f^{ABC} \Phi_f^A \Phi_g^B \Phi_h^C$. One can modify
the superpotential adding the quadratic mass term:
\be
 \label{WM}
 {\cal W}^M \ =\ \frac g{6\sqrt{2}} \epsilon_{fgh} f^{ABC} \Phi_f^A \Phi_g^B 
\Phi_h^C - \frac M2 \Phi_f^A \Phi_f^A
 \ee
The modified scalar field potential 
$U = |\partial {\cal W}^M/\partial \phi_f^A|^2$ turns to zero when the F--terms
vanish
 \be
 \label{Feq0}
\epsilon_{fgh} f^{ABC} \phi_f^A \phi_g^B \ =\ \frac{2\sqrt{2} M}{g} \phi^C_h
\ee
Our matter fields interact also with the gauge fields.
Correspondingly, the $D$--term $\sim  f^{ABC} \phi_f^A \bar \phi_f^B$
is generated. In the vacuum, it also has to vanish.   We have the equation 
system
  \be
\label{DF0}
 \epsilon_{fgh} f^{ABC} \phi_f^A \phi_g^B \ &=&\ C \phi^C_h \ ,
\nonumber \\
   f^{ABC} \phi_f^A \bar \phi_f^B &=& 0 \ ,
  \ee
 $C = 2\sqrt{2}M/g$. Consider first the
$SU(2)$ case. Besides the obvious solution $\phi = 0$, the system (\ref{DF0})
enjoys a unique
up to an overall gauge rotation solution \cite{WitVafa}
 \be
\label{solfi}
\phi_f^A \ =\ \frac 12 C \delta_f^A
 \ee
The appearance of the Higgs average (\ref{solfi}) breaks down the gauge 
invariance completely; all gauge fields and their superpartners acquire mass
of order $M$. As the solution (\ref{solfi}) is unique, the same applies 
to the matter fields irrespectively of whether we are at the vicinity
of classical vacua with $\langle \phi \rangle_{\rm
vac} \sim C$ or  $\langle \phi \rangle_{\rm
vac} = 0$. When mass is large $M \gg E_{\rm char} \sim g^{2/3}$, the 
state (\ref{solfi}) is separated from
the sector $\langle \phi \rangle_{\rm
vac} = 0$ by a high barrier. In the limit $M \to \infty$, this barrier
becomes unpenetrable, and if in the morning we wake up in the sector
with   $\langle \phi \rangle_{\rm
vac} \sim 0$, we are going to stay there also by the end of the day. The
presence of heavy matter fields would not be felt and the dynamics would be the
same as in ${\cal N}=1$ 4--dimensional SYM theory.

On the other hand, when mass is small, the barrier disappears and the new 
vacuum state
overlaps essentially with the conventional vacuum sector. In the limit 
$M \to 0$, the state (\ref{solfi}) goes over into the celebrated localized
supersymmetric vacuum state of the hamiltonian (\ref{H}).

Let us emphasize that the final conclusion that yes, there {\it is}
such a localized supersymmetric state is valid irrespectively of whether we 
are thinking in the language of the SQM system (\ref{Q}, \ref{H}) or in the
language of the associated 4--dimensional field theory. In the former case,
one should speak about a deformation of the hamiltonian (\ref{H}) leaving
only 4 of 16 real supercharges $Q_\alpha$ conserved. Even for non--zero
$M$, the system involves the continuum spectrum associated with the
(4--dimensional) gauge potentials $A_i^A$. For large $M$, the localized
state is well separated from the continuum spectrum states  but, in the limit 
$M \to 0$, it is kind of mixed up with them making the analysis difficult.
Still, the true index, the number of the {\it normalized} vacuum states
does not depend on $M$ and for the $SU(2)$ gauge group, is equal to 1.

If we are thinking in terms of 4--dimensional field theory, the most 
convenient way to treat it is to put it in a finite spatial volume. That
makes the spectrum discrete and just removes all uncertainties connected
with nonzero ``deficit contribution'' in Eq.(\ref{princdef}). If you like, 
going
from quantum mechanics to field theory defined in the box presents a 
convenient infrared regularization making the motion finite {\it and}
preserving supersymmetry \cite{Arc}. It plays the same role as the
compactification $R^4 \to S^4$ for the Dirac operator in gauge field
background. The only difference is that, for the problem (\ref{H}),
such a regularization brings about a lot of (infinitely many) new degrees
of freedom, but as we know since \cite{Wit} how to handle them in case
when the spatial box is small, it is not a real problem.

In the field theory with large mass, we have one extra state at large 
values of Higgs average, and also two conventional vacuum states of 
${\cal N}=1$ SYM theory coming from the region $\phi \sim 0$ where the heavy matter
fields decouple. In this approach, the answer $I_W = 1$ is not obtained
as a difference
$ I_W = I_W^p - I_W^d = \frac 54 - \frac 14 \ $,
but rather as the difference $ I_W = 3 - 2$, with 3 being the Witten index
of the ${\cal N}=4$ SYM field theory while 2 is the Witten index of   ${\cal N}=1$ SYM 
field theory. This reasoning emphasizes again that the separate terms
like $5/4$, $1/4$ or $3,2$ have no particular physical meaning. Only the
total answer $I_W = 1$ is meaningful.

One more remark is in order. As we see, when going from $M = \infty$ to
$M = 0$, the new vacuum state appears in the physical spectrum, and it comes
from infinity of configuration space. This phenomenon is well known. This 
happens e.g. in the models of Wess--Zumino type when the asymptotics of 
superpotential is changed. Various gauge SUSY models involving this phenomenon
have been recently constructed. One example is the ${\cal N}=1$ theory with 
the $G_2$
gauge group where superpotential for matter fields is modified by adding a
cubic term \cite{G2}. Another classic example is ${\cal N}=2$ supersymmetric QCD
\cite{Seiberg}. Some pecularity of ${\cal N}=4$ theory is that we {\it do} not add
here any unusually high power in superpotential, but just a quadratic
mass term (while the cubic term was already there). It was conjectured in 
\cite{Wit} that the Witten index in supersymmetric gauge theories with
{\it non--chiral} matter content (so that the mass terms can be added)
is the same as in the pure SYM theory. We see that it is not true in this 
case. But it is true e.g. in the ${\cal N}=2$ theory where we have only one scalar
field and no cubic superpotential.    

Let us discuss now higher unitary groups. We have to solve again the
equation system (\ref{DF0}), but with an additional requirement: the Higgs
average obtained should break the gauge invariance completely and give
mass to all gauge fields (otherwise, the wave functions would 
smear out along
the flat directions corresponding to the remaining massless fields, and
the state would not be localized.

In mathematical language, that means that we are looking for the triples
$\phi_f = \phi_f^A t^A$ belonging to our Lie algebra ${\mathfrak g}$ which satisfy
the relations (\ref{DF0}) and whose centralizer is trivial (i.e. there is no
such $ g \in {\mathfrak g}$ that $[ g, \phi_f] = 0$
for all $f = 1,2,3$). As was noted in Ref.\cite{WitVafa}, this problem is 
reduced to the mathematical problem of classifying the embeddings
$su(2) \subset {\mathfrak g}$ with trivial centralizer factorized over the 
action of the complexified group $G$. 

Let us prove it. Let $G$ be the complex connected simple Lie group with trivial
center such that ${\mathfrak g}$ is its Lie algebra. The group $G$ acts on
${\mathfrak g}$ faithfully via the adjoint representation. Let $K$ be
a maximal compact subgroup of $G$. Then there exists on ${\mathfrak g}$ 
a unique (up to a positive factor) positive definite Hermitian form 
 such that 
$$ K \ =\ \{g \in G| g^\dagger = g^{-1} \}\ . $$
Let 
$$ P \ =\ \{g \in G| g^\dagger = g \ {\rm and}\ g\ {\rm is \ positive\ 
 definite} \}\ . $$
Then $K \cap P = \{1\}$ and one has the following well-known {\it polar 
decomposition}
 \be
 \label{Kac1}
G \ =\ KP\ .
 \ee
In other words, in any representation, the matrix representing the element of
the complex group $G$ can be written as a product of a unitary and a positively
defined Hermitean matrix.

{\bf Lemma 1}. If $k \in K,\ \ p \in P$, and $pkp^{-1} \in K$, then
$pk = kp$.

{\bf Proof}.
We have:
\be
\label{Kac2}
pk \ =\ k_1p
 \ee
for some $k_1 \in K$. Applying $^\dagger$ to both sides of (\ref{Kac2}),
we get
 \be 
\label{Kac3}
pk_1p^{-1} \ =\ k
 \ee
Comparing (\ref{Kac2}) and (\ref{Kac3}), we get
  \be 
\label{Kac4}
p^2kp^{-2} \ =\ k
 \ee
Thus, $p^2$ commutes with $k$, but since $p$ is positive definite Hermitean,
we deduce that $p$ commutes with $k$ as well.

{\bf Corollary 1}. If ${\mathfrak a}_1$ and  ${\mathfrak a}_2$ are two real
subalgebras of $Lie\ K \subset {\mathfrak g}$ isomorphic to $su(2)$ such that
they are conjugate by an element $g \in G$, then they are conjugate by an 
element of $K$.

{\bf Proof}. We write $g = kp$, where $k \in K, \ p\in P$. Then 
$$ (Ad\ p) {\mathfrak a}_1 \ =\ (Ad\ k^{-1}) {\mathfrak a}_2 
\ \subset Lie \ K $$
Denoting by $A_i$ the subgroup of $K$ whose Lie algebras are ${\mathfrak a}_i$,
we deduce:
$$ (Ad\ p) A_1 \ =\ (Ad\ k^{-1}) A_2 
\ \subset  \ K\ . $$
It follows from Lemma 1 that $(Ad\ p) A_1 \ = A_1$, hence ${\mathfrak a}_1$
and ${\mathfrak a}_2$ are conjugate by $k$.

{\bf Theorem 1}. Triples of elements $(T_1, T_2, T_3)$ of ${\mathfrak g}$
satisfying the equations

{\it (i)} $[T_j, T_k] = i \epsilon_{jkl} T_l$

{\it (ii)} $[T_j, T_j^\dagger] = 0$

are conjugate by $G$ if and only if they are conjugate by $K$.

{\bf Proof}. Condition {\it (i)} means that $T_j$ form a basis of $su(2) 
\subset {\mathfrak g}$. In view of Corollary 1, it suffices to show that 
conditions {\it (i)} and {\it (ii)} imply that $T_j = T_j^\dagger$. 

We may assume that $T_j \subset Lie\ K$ (since the maximal compact subgroups
of $G$ are conjugate). Thus, we have a homomorphism $\phi :\ sl(2) \to 
{\mathfrak g} $ such that $\phi[su(2)] \subset K$. Hence it suffices to show
that equations {\it (i)} and {\it (ii)} on three elements $T_j$ of $sl(2)$
imply that $T_j \in su(2)$. 

Let $\sigma_1$, $\sigma_2$, and $\sigma_3$ be an orthonormal basis of 
$su(2)$. We have for some  $g \in SL(2)$ : 
 $$ g T_j g^{-1} \ =\ \sigma_j\ ,\ \ \ \ \ \ j = 1,2,3 \ .$$
Writing  $g = kp$ (polar decomposition) and replacing $\sigma_j$ by
$k^{-1}\sigma_j k$, equation {\it (ii)} gives: 
  \be
 \label{Kac5}
\sum_j p^2 (\sigma_j p^{-2} \sigma_j ) p^2 \ =\ \sum_j \sigma_j p^2 \sigma_j\ ,
 \ee
where $p^2$ is a positive definite Hermitean matrix. Choosing a suitable 
basis, we may  assume that $\sigma_j$ are the Pauli matrices. A direct 
calculation shows that (\ref{Kac5}) implies that $p =1$.

When $G = SU(n)$, there is only one such embedding (we will prove it 
rigourously in the next section). It is sufficient to write down the
generators of $SU(2)$ in the representation with the spin $j = (n-1)/2$ and
treat them as the elements of the $su(n)$ algebra in the fundamental 
representation. For example, for $su(3)$, the non--trivial triple is
 \be
 \label{solsu3}
\phi_1 + i\phi_2 \ =\ \left( 
\begin{array}{ccc}
0&1&0 \\ 0&0&1 \\ 0&0&0 \end{array} \right),\ \ 
\phi_1 - i\phi_2 \ =\ \left( 
\begin{array}{ccc}
0&0&0 \\ 1&0&0 \\ 0&1&0 \end{array} \right),\ \ 
\phi_3 \ =\ \left( 
\begin{array}{ccc}
1&0&0 \\ 0&0&0 \\ 0&0&-1 \end{array} \right)
 \ee
The existence and uniqueness of the solution means that the ${\cal N}=4$ theory 
with $SU(n)$ gauge group has one and only one normalized vacuum state in 
agreement with (\ref{sunres}).

\section{Distinguished $sl(2)$ subalgebras in   simple 
Lie algebras.}
\setcounter{equation}0

The problem is reduced to finding all the solutions of the equation system
(\ref{DF0}) for an arbitrary gauge group. It is not a trivial problem but,
fortunately, its solution can be easily derived from related problems
that  have actually already been solved by mathematicians.

Let us first introduce some basic notations and definitions.
Let  ${\mathfrak g}$ be a complex simple  Lie algebra and $G$ be the
corresponding complexified group. Choose a Cartan subalgebra 
${\mathfrak h}$ in ${\mathfrak g}$ . A convenient choice of basis
in   ${\mathfrak g}$ is a union of the basis  of 
${\mathfrak h}$  and  the {\it root vectors} 
$e_{{\bf \alpha} }$: $[h, e_{{\bf \alpha} }] = 
\alpha (h) e_{{\bf \alpha} }$ for any $h \in {\mathfrak h}$, 
 $\alpha $ (the linear forms on the Cartan subalgebra) being 
the {\it roots}.
\footnote{In other notation, $[h_i, e_{{\bf \alpha} }] = 
\alpha_i e_{{\bf \alpha} }$ for a particular basis $h_i$ in ${\mathfrak h}$.}
 For any root $\alpha $, $-\alpha $ is also a root,
and the whole set of roots $\Delta$ can be decomposed into a set of positive
roots $\Delta_+$ and a set of negative roots $\Delta_-$. If $- \alpha \in 
\Delta_-$, we will use the notation $f_{\bf \alpha}$ for $e_{ - {\bf \alpha}}$.
The commutator $[e_{\bf \alpha}, f_{\bf \alpha}] \equiv \alpha^\lor$ lies in
the Cartan subalgebra. With the standard choice of normalization for the root
vectors, $[\alpha^\lor, e_\alpha] = 2e_\alpha$ and 
$[\alpha^\lor, f_\alpha] = -2f_\alpha$, $\alpha^\lor$ is called the 
{\it coroot}. 
For any coroot $\alpha^\lor$, the identity
 \be
\label{unity}
\exp\{2\pi  i \alpha^\lor \}\ =\ 1 \in G
 \ee
holds.
For any $\alpha, \beta \in \Delta$ with $\alpha + \beta \neq 0$,
$[e_\alpha, e_\beta]$ is proportional to  $ e_{\alpha + \beta}$ with non--zero
coefficient if $\alpha + \beta \in \Delta$, and
$[e_\alpha, e_\beta] = 0$ otherwise. A set of $r$ {\it simple roots} 
$\alpha^{(i)}$ and the corresponding simple root vectors $e_i, f_i$ can be 
chosen
so that all other root vectors are obtained from  $e_i, f_i$ by a number
of subsequent commutations. The corresponding coroots 
$\alpha_{(i)}^\lor \equiv
h_i$ present a convenient basis in the Cartan subalgebra. The set 
$\{e_i, f_i, h_i\}$ is called the {\it Chevalley generators}. An element 
$\omega_i \in {\mathfrak h}$ 
commuting with all but one pair of simple root vectors, so that
 $$[\omega_i, e_j]
= \delta_{ij} e_j, \ \ \ [\omega_i, f_j]
= - \delta_{ij} f_j \ ,$$
is called  {\it fundamental coweight}.

An element $ x$ of a Lie algebra  ${\mathfrak g}$ is called
{\it nilpotent} (resp. {\it semisimple}) if $ ad\  x \equiv [x, $ is a 
nilpotent (resp. diagonalizable) operator on  ${\mathfrak g}$. A subalgebra
of  ${\mathfrak g}$ is called {\it reductive} if it is a direct sum of  simple
subalgebras and a torus (i.e. commutative subalgebra consisting of semisimple
elements). 

The following result is due to Morozov, Jacobson,  Dynkin, and Kostant. 
Its proof can be
found in \cite{Dyn,Jacob,Kost}.

{\bf Theorem 2}. Let $e$ be a non--zero nilpotent element of a simple complex
Lie algebra ${\mathfrak g}$. Then
 
{\it (a)} There exist elements $h,f \in {\mathfrak g}$ such that 
 \be 
\label{sl2}
 [h, e] = e,\ \ \ \ [h, f] = -f,\ \ \ [e,f] = h\ ,
 \ee
i.e. ${\bf C}e + {\bf C}h + {\bf C}f$ is isomorphic to the 3--dimensional
simple algebra $sl(2,{\bf C})$. In this case, the element $f$ is nilpotent
and element $h$ is semisimple with integer or half-integer  eigenvalues.

{\it (b)} If  ${\bf C}e + {\bf C}h' + {\bf C}f'$ is another 3--dimensional
simple algebra containing $e$, then there exists $g \in G$ such that 
$g(e) = e, \  g(h) = h', \ g(f) = f'$.

{\it (c)} There is a bijective correspondence between conjugacy classes of non--zero nilpotent elements of  ${\mathfrak g}$ and conjugacy classes of 
3--dimensional simple subalgebras of  ${\mathfrak g}$. 

\vspace{.3cm}

It follows from Theorem 2a that one has the eigenspace decomposition with
respect to $ad \ h$ : 
\be
\label{gradj}
 {\mathfrak g} \ =\ \oplus_{j \in {\bf Z}/2}  \ {\mathfrak g}_j,\ \ \ 
[ {\mathfrak g}_i,  {\mathfrak g}_j] \subset  {\mathfrak g}_{i+j}, \ \ 
e \in  {\mathfrak g}_1,\ \ f \in  {\mathfrak g}_{-1}
 \ee
In other words, $[h,x] = jx$ if $x \in   {\mathfrak g}_j$. Let 
$ {\mathfrak g}_+ =  \oplus_{j > 0}  \ {\mathfrak g}_j$,
$ {\mathfrak g}_- =  \oplus_{j < 0}  \ {\mathfrak g}_j$. The proof of the 
following result may be found in \cite{BC}.

{\bf Theorem 3}. 

{\it (a)} The centralizer of $e$ (resp. $f$) in $ {\mathfrak g}$ is a sum
of a reductive subalgebra ${\mathfrak m}^+$ of ${\mathfrak g}_0$ and a 
subalgebra of ${\mathfrak g}_+$ (resp. ${\mathfrak g}_-$) consisting of
nilpotent elements (of ${\mathfrak g}$).

{\it (b)} $[{\mathfrak m}^+, e] = {\mathfrak g}_1$, 
$[{\mathfrak m}^-, f] = {\mathfrak g}_{-1}$. In other words, the 
$M^+$--orbit of $e$ (resp. $M^-$--orbit of $f$) in  ${\mathfrak g}_1$ 
( resp.  ${\mathfrak g}_{-1}$)
is open and dense.

{\it (c)} $ ad\ e$: $ {\mathfrak g}_0 \to  {\mathfrak g}_1$ and
 $ ad\ f$: $ {\mathfrak g}_0 \to  {\mathfrak g}_{-1}$ are surjective linear
maps. In particular, dim $ {\mathfrak g}_0 \geq  {\rm dim} \ {\mathfrak g}_1$.

\vspace{.2cm}

A nilpotent element $e$ is called {\it distinguished} if 
$ {\mathfrak m}^+ = 0$. The following result is straightforward. Its proof
 may be found in \cite{Dyn,BC}.

{\bf Theorem 4}. A nilpotent element $e$ is distinguished if one of the 
following equivalent properties holds:

{\it (i)} The centralizer of $e$ in ${\mathfrak g}$ lies in ${\mathfrak g}_+$,
i.e. it consists of nilpotent elements. 

{\it (ii)} dim ${\mathfrak g}_0$ = dim ${\mathfrak g}_1$.

{\it (iii)} dim ${\mathfrak g}_0$ = dim ${\mathfrak g}_{-1}$.
 
\vspace{.2cm}

 A 3--dimensional simple subalgebra of ${\mathfrak g}$ is called 
{\it distinguished} if its (unique up to conjugacy) non--zero nilpotent
element is distinguished.

{\bf Theorem 5}. A 3--dimensional simple subalgebra ${\mathfrak a}$ of
  ${\mathfrak g}$ is distinguished if and only if its centralizer in
${\mathfrak g}$ is zero.

{\bf Proof}. Let  ${\mathfrak a}$ be a distingushed subalgebra of 
${\mathfrak g}$ and let $e$ be the corresponding (distingushed) 
nilpotent element. But the centralizer $C({\mathfrak a})$ of
${\mathfrak a}$ in ${\mathfrak g}$ is a reductive subalgebra which lies in
the centralizer of $e$, which consists of nilpotent elements due to Theorem
3a, hence  $C({\mathfrak a}) = 0$.

Conversely, let  ${\mathfrak a} = {\bf C}e + {\bf C}h + {\bf C}f$ be a
3--dimensional simple subalgebra of ${\mathfrak g}$ with zero centralizer.
Then with respect to the adjoint representation of ${\mathfrak a}$ in
${\mathfrak g}$, ${\mathfrak g}$ decomposes into a direct sum of non--trivial
irreducible submodules $V_i$ such that dim$(V_i \cap {\mathfrak g}_j) = 1$
if $|j| \leq s $ and $s - j \in {\bf Z}$, and
 dim$(V_i \cap {\mathfrak g}_j) = 0$ otherwise [see (\ref{gradj})]. It follows
that  dim ${\mathfrak g}_0$ = dim ${\mathfrak g}_1$  = 
dim ${\mathfrak g}_{-1}$, hence $e$ is 
distinguished by Theorem 4.

\vspace{.2cm}

Choose Chevalley generators $e_i, h_i, f_i$ ($i = 1,\ldots,r$) of ${\mathfrak 
g}$. The Dynkin diagram with dots on  some of its nodes is called
{\it marked}. Such a marking defines a ${\bf Z}$--gradation ${\mathfrak 
g} = \oplus_j \ {\mathfrak g}_j$ if we let deg $e_i$ = - deg $f_i$ = 1 if the
i-th node is marked, and deg $e_i$ = deg $f_i$ = 0  otherwise. A marking is 
called
{\it distinguished} if
  \be
 \label{dimdim}
{\rm dim} \ {\mathfrak g}_0 \ =\ {\rm dim}\ {\mathfrak g}_1
 \ee
According to Dynkin  \cite{Dyn} (see also \cite{BC}), one has a bijective 
correspondence between conjugacy classes of distinguished nilpotent elements
of ${\mathfrak g}$ and distinguished markings of the Dynkin diagram of
${\mathfrak g}$. Namely, given a distingushed nilpotent element $e$ of 
${\mathfrak g}$, we construct a {\bf Z}--gradation (\ref{gradj}) of 
${\mathfrak g}$
by ad $h$ and choose a set of positive roots $\Delta_+$ such that
$[h, e_\alpha] \equiv \frac 12 \alpha(h)e_\alpha $ if $\alpha \in \Delta_+$.
It turnes out that $\alpha(h) = 0$ or 2 if $\alpha$ is a simple root, hence
we get a {\bf Z}--gradation of ${\mathfrak g}$ corresponding to a marked
Dynkin diagram. Conversely, given such a gradation, we pick $e \in {\mathfrak 
g}_1$ such that $[{\mathfrak g}_0, e] = {\mathfrak g}_1$ and take $h \in
{\mathfrak g}_0$ such that ad $h$ defines this gradation. Due to
(\ref{dimdim}), there exists a unique $f \in {\mathfrak g}_{-1} $ such that
$[e, f] = h$ giving a 3--dimensional simple subalgebra with zero centralizer.

{\bf Example 1}. Look again at Eq.(\ref{solsu3}) defining the distingushed
$sl(2)$ subalgebra in the $sl(3)$ algebra. We have
 \be
 \label{tripsu3}
\phi_1 + i\phi_2 = e_\alpha + e_\beta \equiv e\ , \ \ 
\phi_1 - i\phi_2 = f_\alpha + f_\beta \equiv f\ , \ \
\phi_3 = \alpha^\lor + \beta^\lor \equiv h\ , 
\ee 
  where $\alpha$ and $\beta$ are two simple roots. The gradation defined by $h$
involves: {\it (i)} ${\mathfrak g}_0$ which coincides in this case with 
the Cartan subalgebra; {\it (ii)} ${\mathfrak g}_1$ (resp. ${\mathfrak g}_{-1}$) with the basis $e_\alpha$, $e_\beta$ ( resp. $f_\alpha$, $f_\beta$); and
{\it (iii)} ${\mathfrak g}_2$ (resp. ${\mathfrak g}_{-2}$) with the basis 
$e_{\alpha+\beta}$ ( resp. $f_{\alpha+\beta}$). Obviously, the condition
(\ref{dimdim}) is satisfied. 

{\bf Example 2}. Consider an arbitrary simple Lie algebra ${\mathfrak g}$.
Choose an element $\rho \in {\mathfrak h}$ such that $[\rho, e_i] = e_i$
for all positive simple roots $e_i$. The element $\rho$ defines the canonical
gradation of the Lie algebra (corresponding to the Dynkin diagram
with all nodes marked) such that ${\mathfrak g}_0$ is the Cartan 
subalgebra; the basis of  ${\mathfrak g}_1$ ( of ${\mathfrak g}_{-1}$ ) is the
system of positive (negative) simple root vectors;
  ${\mathfrak g}_2$ (resp.  ${\mathfrak g}_{-2}$ ) is spanned by the root 
vectors of level 2 (resp. $-2$), etc. Obviously, ${\rm dim}\ {\mathfrak g}_0 
\ =\ {\rm dim}\ {\mathfrak g}_{1} \ =\  {\rm dim}\ {\mathfrak g}_{-
1} = r$. 

Let $\rho = \sum_{i=1}^r b_i h_i$ where $h_i$ are the simple coroots;
then $b_i$ are positive numbers.
The triple 
\be
\label{efro}
e \ =\ \sum_{i=1}^r e_i  \sqrt{b_i},\ \ f \ =\ \sum_{i=1}^r f_i  
\sqrt{b_i},\ \ h =\ \rho
  \ee
form the distinguished $sl(2)$ subalgebra. In the $sl(n)$ case, Eq.(\ref{efro})
is reduced to the known distinguished embedding described at the end of the
previous section.
 
We see thereby that, for any   ${\mathfrak g}$, a solution of the equations
(\ref{DF0}) exists which provides us with at least one normalized vacuum state
for any gauge group $G$. For most of simple groups, however, the solution is ]
not unique.
The simplest group involving more that one supersymmetric vacuum is the group
$G_2$.

\begin{figure}
\grpicture{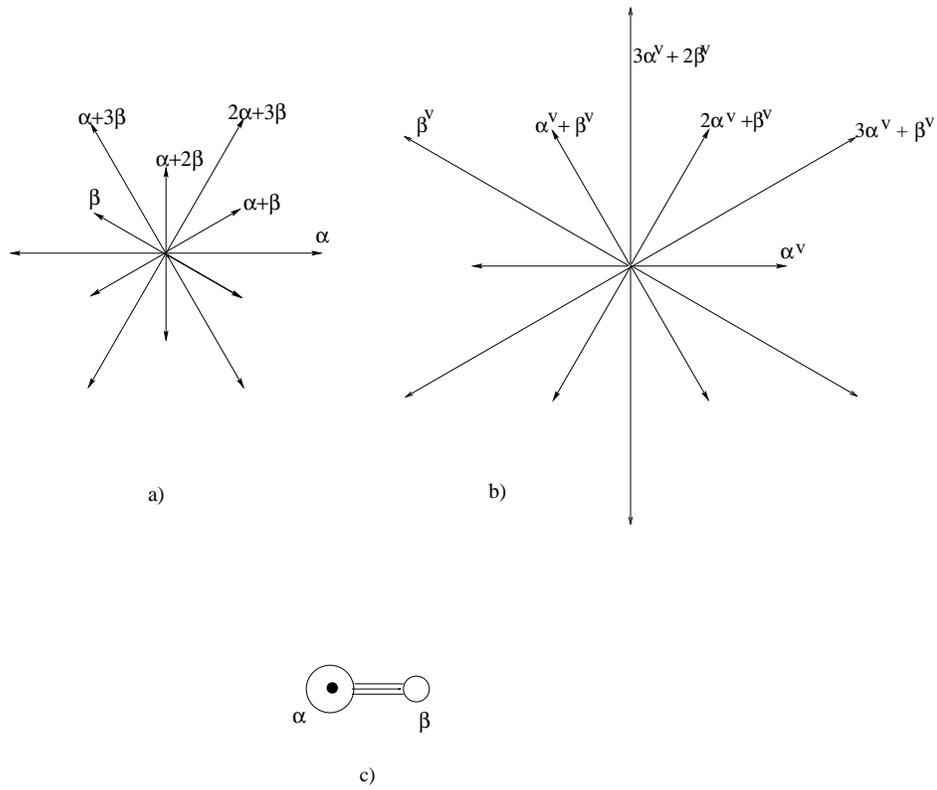}
\caption{{\it a)}  root system, {\it b)} coroot system, 
and {\it c)} non--trivial distinguished
marked Dynkin diagram for  the group $G_2$.}
\label{G2}
\end{figure}

{\bf Example 3}. The system of  roots of $G_2$ is depicted in Fig.1a,
the system of corresponding coroots
\footnote{The coroot corresponding to a long coroot is short and 
the coroot corresponding to a short coroot is long.} in Fig.1b, 
and the Dynkin diagram is drawn in Fig. 1c. We have put the dot on the
long simple root $\alpha$, but not on the short root $\beta$ (Here and in the
following short roots will be denoted by smaller circles). Thereby, a 
certain
non--trivial marking of the Dynkin diagram is defined. The corresponding
{\bf Z}--gradation involves:
\be
 \label{gradG2}
  {\mathfrak g}_0 \ &{\rm with \ the \ basis}& \  \alpha^\lor, \beta^\lor,
e_\beta, \ {\rm  and} \ f_\beta \nonumber \\  
  {\mathfrak g}_1 \ & {\rm with\  the\  basis}&\ 
e_\alpha, e_{\alpha+\beta}, 
 e_{\alpha+2\beta} , \  {\rm and}\ 
 e_{\alpha+3\beta} ; \nonumber \\
 {\mathfrak g}_{-1} \ &{\rm with \ the \ basis}& \ 
f_\alpha, f_{\alpha+\beta} , 
 f_{\alpha+2\beta} , \ {\rm and}\ 
 f_{\alpha+3\beta} ; \nonumber \\
   {\mathfrak g}_2 \ &{\rm with \ the \ basis}&  \   e_{2\alpha + 3\beta} ;\ \ 
  \nonumber \\
 {\mathfrak g}_{-2} \ &{\rm with\  the \ basis}& \   f_{2\alpha + 3\beta}
 \ee
 The condition (\ref{dimdim}) is satisfied, and hence the marking in Fig. 1b
is distinguished.

Let us construct now the corresponding distingushed $sl(2)$ subalgebra, the
distinguished triple $(e,f,h)$. First, let us find the required element
 of the Cartan subalgebra 
 \be
\label{hG2}
h:\ \ \ [h, e_\alpha] = e_\alpha,\ \ \ \ \ [h,e_\beta] = 0
  \ee
so that the gradation (\ref{gradG2}) is realized
by the action of $ad\ h$. This is just the fundamental coweight corresponding
to the node $\alpha$ of the Dynkin diagram. In our case, $h = (2\alpha
+ 3\beta)^\lor = 2\alpha^\lor + \beta^\lor $. To find explicitly the elements 
$e,f$ of the triple, write
$e$ as a generic element of ${\mathfrak g}_1$
$$ e \ =\ a_1 e_\alpha + a_2 e_{\alpha+\beta} + a_3 e_{\alpha+2\beta}
+ a_4 e_{\alpha+3\beta}\ ,$$
choose
$$ f \in {\mathfrak g}_{-1} \ =\ 
\bar a_1 f_\alpha + 
\bar a_2 f_{\alpha+\beta} + \bar a_3 f_{\alpha+2\beta}
+ \bar a_4 f_{\alpha+3\beta}\ ,$$
and impose the requirement $[e,f] = h$.
Substituting here the standard commutators
   \be
[e_{\alpha + \beta} , f_{\alpha + \beta}] \ = \ (\alpha + \beta)^\lor \ =\ 
3\alpha^\lor + \beta^\lor  \nonumber  
 \ee
\vspace{-1cm}
\be
[  e_{\alpha + 2\beta} , 
f_{\alpha + 2\beta} ] \ = \ 
(\alpha + 2\beta)^\lor \ =\ 
3\alpha^\lor + 2\beta^\lor \nonumber
 \ee
\vspace{-1cm}
\be
[e_{\alpha + 3\beta} , 
f_{\alpha + 3\beta}] \ = \ 
(\alpha + 3\beta)^\lor \ =\ 
\alpha^\lor + \beta^\lor \nonumber
 \ee
\vspace{-1cm}
 \be
[e_{\alpha + \beta} , 
f_{\alpha}] \ 
=\ \frac 12 
[e_{\alpha + 2\beta}, 
f_{\alpha + \beta}] \ =\ 
[e_{\alpha + 3\beta}, 
f_{\alpha + 2\beta}] \ =\
-e_\beta\ , 
 \ee
 we obtain 3 equations for 4 complex
parameters $a_i$:
 \be
\label{eqaG2}
 |a_1|^2 +  3|a_2|^2 + 3|a_3|^2 + |a_4|^2 \ &=& 2 \nonumber \\
 |a_2|^2 + 2|a_3|^2 + |a_4|^2 \ &=& 1 \nonumber \\
a_2 \bar a_1  + 2 a_3 \bar a_2 + a_4 \bar a_3 &=& 0
 \ee
Different solutions to this equation system are related to
each other by conjugation.  The convenient choice is $a_1 = a_4 = 1,\ 
a_2 = a_3 = 0$ which gives the triple
 \be
\label{tripG2}
 e \ =\  e_\alpha + e_{\alpha + 3\beta},\ \ \  
f \ =\  f_\alpha + f_{\alpha + 3\beta},\ \ \ h = 2\alpha^\lor + \beta^\lor
 \ee
 The distinguished $sl(2)$ subalgebra of the Lie algebra of $G_2$ with the
basis (\ref{tripG2}) 
{\it is} not equivalent by conjugation to the universal subalgebra 
(\ref{efro}).
Therefore, the ${\cal N}=4$ supersymmetric Yang--Mills quantum mechanics with the 
$G_2$ gauge
group has two different normalized supersymmetric vacuum states.

\begin{figure}
    \begin{center}
        \epsfxsize=200pt
        \epsfysize=0pt
        \vspace{-5mm}
        \parbox{\epsfxsize}{\epsffile{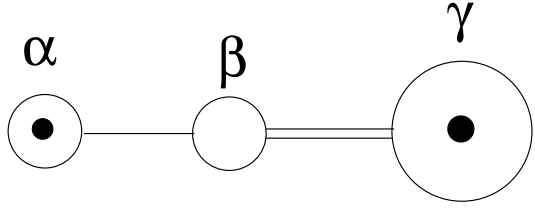}}
        \vspace{5mm}
    \end{center}
\caption{ Non--trivial distinguished marked Dynkin diagram for 
the group $Sp(6)$.}
\label{Sp6}
\end{figure}

{\bf Example 4}. Consider the marked 
 Dynkin diagram for the $Sp(6)$ group  drawn in Fig. \ref{Sp6}. 
 The corresponding
{\bf Z}--gradation involves:
\be
 \label{gradSp6}
  {\mathfrak g}_0 \ &{\rm with \ the \ basis}&\  \alpha^\lor, \ \beta^\lor,
\ \gamma^\lor,  \ e_\beta,  \ f_\beta\ , \nonumber \\  
  {\mathfrak g}_1 \ &{\rm with \ the \ basis}& \ 
e_\alpha, \ e_{\alpha + \beta}, \ e_\gamma, \ e_{\beta+\gamma}, 
e_{2\beta+\gamma} \ , \nonumber \\  
{\mathfrak g}_2 \ &{\rm with \ the \ basis}& \   e_{\alpha + \beta + \gamma},
\  e_{\alpha + 2\beta + \gamma}\ ,
 \nonumber \\
  {\mathfrak g}_{3} \ &{\rm with \ the \ basis}&\ 
  e_{2\alpha + 2\beta + \gamma } \ ,
 \ee
and the subalgebras $ {\mathfrak g}_{-1}$,  $ {\mathfrak g}_{-2}$,  
$ {\mathfrak g}_{-3}$ spanned by the corresponding negative root vectors.
 The condition (\ref{dimdim}) is satisfied, and hence the marking in Fig. 
\ref{Sp6} is distinguished.

The element $h$ of our
distinguished triple realizing the gradation (\ref{gradSp6}) is the sum of the 
fundamental coweights corresponding
to the marked nodes: 
 \be
\label{hSp6}
h \ =\ \omega_\alpha  + \omega_\gamma
\ =\ \frac 32 \alpha^\lor + 2\beta^\lor + \frac 52 \gamma^\lor
 \ee
  The elements $e \in {\mathfrak g}_1$  and  $f \in {\mathfrak g}_{-1}$
can be found in the same way as in the previous example. We have a system
of 4 equations for 5 complex coefficients. One of the solutions has the
form
 \be
\label{efSp6}
 e \ &=&\  \sqrt{\frac 32} \ e_\alpha +   \frac 1{\sqrt{2}} \ e_\gamma
+  \sqrt{2} \ e_{2\beta + \gamma}   \nonumber \\
 f \ &=&\  \sqrt{\frac 32} \ f_\alpha +   \frac 1{\sqrt{2}} \ f_\gamma
+  \sqrt{2} \ f_{2\beta + \gamma}  \ .
 \ee
 All other solutions of this equation system are equivalent to
Eq(\ref{efSp6}) by conjugation. The triples (\ref{efSp6}) and 
(\ref{efro}) present two inequivalent by conjugation  
 distinguished $sl(2)$ subalgebras 
of the Lie algebra of $Sp(6)$. This gives  
 two different normalized supersymmetric vacuum states.

\begin{figure}
    \begin{center}
        \epsfxsize=200pt
        \epsfysize=0pt
        \vspace{-5mm}
        \parbox{\epsfxsize}{\epsffile{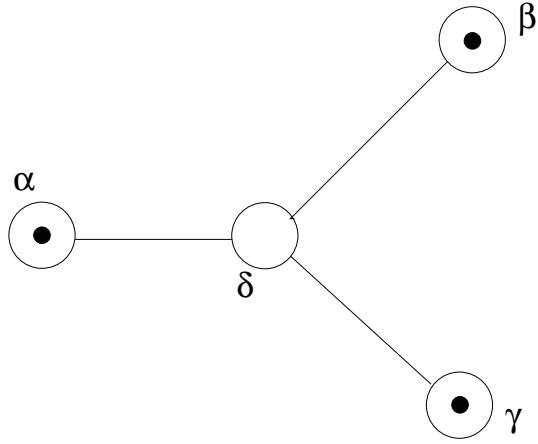}}
        \vspace{5mm}
    \end{center}
\caption{ Non--trivial distinguished marked Dynkin diagram for 
the group $SO(8)$.}
\label{O8}
\end{figure}

{\bf Example 5}. Consider the marked 
 Dynkin diagram for the $SO(8)$ group  drawn in Fig. \ref{O8}. 
 The corresponding
{\bf Z}--gradation involves:
\be
 \label{gradO8}
  {\mathfrak g}_0 \ &{\rm with \ the \ basis}&\  \alpha^\lor, \ \beta^\lor,
\ \gamma^\lor, \ \delta^\lor, \ e_\delta,  \ f_\delta \nonumber \\  
  {\mathfrak g}_1 \ &{\rm with \ the \ basis}& \ 
e_\alpha, \ e_\beta, \ e_\gamma, \ e_{\alpha+\delta},  \ e_{\beta+\delta},
\  e_{\gamma+\delta}, \nonumber \\  
{\mathfrak g}_2 \ &{\rm with \ the \ basis}& \   e_{\alpha + \beta + \delta},
\  e_{\gamma + \beta + \delta}, \ e_{\alpha + \gamma + \delta},
 \nonumber \\
  {\mathfrak g}_{3} \ &{\rm with \ the \ basis}&\ 
  e_{\alpha + \beta + \gamma + \delta}, \ 
e_{\alpha+\beta+\gamma + 2\delta}\ \ ,
 \ee
and the corresponding subalgebras 
$ {\mathfrak g}_{-1}$,  $ {\mathfrak g}_{-2}$,  
$ {\mathfrak g}_{-3}$.
 We have dim($ {\mathfrak g}_0$) = dim($ {\mathfrak g}_{\pm 1}$) = 5, and
 hence the marking in Fig. \ref{O8} is distinguished.

The distingushed triple can be constructed  along the same lines as in the 
previous example. It is  (up to a conjugation):
 \be
\label{efO8}
h \ &=&\  2(\alpha^\lor + \beta^\lor + \gamma^\lor) + 3\delta^\lor\ , 
\nonumber \\
 e \ &=&\  e_\alpha + e^{i\pi/3} e_\beta +  e^{2i\pi/3} e_\gamma
+  e_{\alpha + \delta} + e^{-i\pi/3} e_{\beta + \delta} + 
 e^{-2i\pi/3} e_{\gamma + \delta}  \nonumber \\
   f \ &=&\  f_\alpha + e^{-i\pi/3} f_\beta +  e^{-2i\pi/3} f_\gamma
+  f_{\alpha + \delta} + e^{i\pi/3} f_{\beta + \delta} + 
 e^{2i\pi/3} f_{\gamma + \delta} \ .
 \ee
This triple together with the universal triple (\ref{efro}) gives us
 two different normalized supersymmetric vacuum states.

\vspace{.2cm}

The full classification of distinguished markings for all algebras was done
in Refs.\cite{Dyn,BC}. Translating it into our physical language gives 
immediately the following result:

{\bf Theorem 6}. The number of inequivalent  by conjugation solutions
of Eq.(\ref{DF0}) and hence the number $\#_{\rm vac}[G]$ of different 
normalized 
supersymmetric vacua in the theory (\ref{H}) with gauge group $G$ is

{\it (a)} $\#_{\rm vac}[SU(n)] = 1$.

{\it (b)} $\#_{\rm vac}[Sp(2r)] $ 
 coincides with the number of partitions of $r$ into distinct parts.
\footnote{ For $r =1$ or $r=2$, we have only one universal solution 
(\ref{efro}). For $r = 3 = 3  = 2+1$, we have two inequivalent solutions,
for $r=6 = 6  = 5+1 = 4+2 = 1+2+3$, we have four solutions, etc. }

{\it (c)}  $\#_{\rm vac}[SO(n)] $   coincides with the number of partitions 
of $n$ into distinct odd \\ parts.
\footnote{The second solution appears starting from $n = 8 = 7+1 = 5+3$ and
$n = 9 = 9 = 1+3+5$.}

{\it (d)}  $\#_{\rm vac}[G_2]  = 2$,  $\#_{\rm vac}[F_4] = 4 $, 
 $\#_{\rm vac}[E_6]  = 3$,  $\#_{\rm vac}[E_7] = 6$, and
 $\#_{\rm vac}[E_8]  = 11$. 

\vspace{.3cm}

 A nilpotent element of $sp(n)$ [resp. $so(n)$]
is distinguished iff, viewed as an element of $gl(n)$ it can be conjugated
to a Jordan
form with distinct even (resp. odd) sizes of Jordan blocks, and this
Jordan form completely determines the conjugacy class. Denote by
$s$ the number of these blocks; $s$ just coincides with the number of nonzero
distinct parts in the partition of $r = n/2$ [resp. $n$].  

Let $\bar G = G/Z$ ($Z$ --- center of $G$) be the adjoint group. 
Let $sl(2) = {\bf C} e + {\bf C} h + {\bf C} f \ \subset
{\mathfrak g}$ and let  $\bar G_{e,h,f}$ be the centralizer of this 
$sl(2)$ in $\bar G$. By a theorem of Kostant \cite{Kost},  
$\bar G_{e,h,f}$ is the 
maximal reductive subgroup of $\bar G_e$ (the 
centralizer of $e$ in $\bar G$). 
 The  $sl(2)$ is distinguished if and only if the
group $\bar G_{e,h,f}$
 is a finite group isomorphic to the group of components of $\bar G_e$. 
The groups $\bar G_{e,h,f}$ are always 
trivial  
in the $sl(n)$ case and are isomorphic to ${\bf Z}_2^{s-1}$ in the
$sp(2r)$ and $so(2r+1)$ cases  and to  ${\bf Z}_2^{s-2}$ in the $so(2r)$ case,
 where $s$ is the number of Jordan blocks of the nilpotent elements 
(see above). The nilpotent element
$e$  in the universal distinguished triple (\ref{efro}) consists of just one 
block for $sp(2r)$ and $so(2r+1)$ and of two blocks
for $so(2r)$, and
$\bar G_{e,h,f}$ is always trivial.

Let us illustrate it in the  $Sp(6)$ example. $Sp(6)$ is a 
subgroup  of $SU(6)$ leaving invariant the form $\psi_\alpha C^{\alpha\beta}
\chi_\beta$  where $\psi_\alpha$ and $\chi_\alpha$ are some  {\bf 6}--plets of
$SU(6)$ and the antisymmetric symplectic matrix $C$ can be
chosen in the form
 \be
 \label{Csympl}
C \ =\ \left( \begin{array}{cccccc}
0&0&0&0&0&1 \\  0&0&0&0&1&0 \\ 0&0&0&1&0&0 \\  0&0&-1&0&0&0 \\
0&-1&0&0&0&0 \\ -1&0&0&0&0&0  \end{array}  \right)\ \ .
  \ee
 Then the coroots of $Sp(6)$ are represented by the diagonal
matrices $6 \times 6$:
 \be
\label{corootSp}
 \alpha^\lor &=& {\rm diag}(1,-1,0,\ 0,1,-1), \nonumber \\ 
 \beta^\lor &=& {\rm diag}(0,1,-1,\ 1,-1,0), \nonumber \\  
\gamma^\lor &=& {\rm diag}(0,0,1,\ -1,0,0)\ .
  \ee
The triple (\ref{hSp6}), (\ref{efSp6}) acquires the form
  \be
\label{tripSpmat}
h = \frac 12 {\rm diag} (3,1,1,-1,-1,-3),\ \ 
e = \frac 1 {\sqrt{2}} \left(  \begin{array}{cccccc}
0 & \sqrt{3} & 0 &  0 & 0 & 0 \\
0 & 0 & 0 &  0 & 2 & 0 \\
0 & 0 & 0 &  1 & 0 & 0 \\
0 & 0 & 0 &  0 & 0 & 0 \\
0 & 0 & 0 &  0 & 0 & -\sqrt{3} \\
0 & 0 & 0 &  0 & 0 & 0 
\end{array}    \right),\ \ f =  e^T
  \ee
Indeed, we see that the  nilpotent element $e$ viewed as a
$6 \times 6$ matrix involves two Jordan blocks: 
 \be
\label{Jordan}
J_1 = 
\left(  \begin{array}{cc}
 0 & 1 \\
0 & 0         \end{array}    \right),\ \ \ 
J_2 = 
\left(  \begin{array}{cccc}
 0 & \sqrt{3} & 0 & 0 \\
0 & 0 & 2 & 0 \\ 
 0 &  0 & 0 & -\sqrt{3} \\
0 & 0 & 0 & 0
\end{array}    \right)\ .
 \ee
The block $J_1$ is formed be the ``center'' of the matrix (the columns and
rows 3,4) and the block $J_2$ --- by its ``periphery''. The triple 
(\ref{tripSpmat}) corresponds to the partition $r = 3 = 2 + 1$. $2 \times 2
= 4$ and $ 2 \times 1 = 2$ are the dimensions of the Jordan blocks in 
(\ref{Jordan}). The centralizer of the triple (\ref{tripSpmat}) in
$Sp(6)/{\bf Z}_2$ is ${\bf Z}_2$ whose nontrivial element is
diag$(1,1,-1,-1,1,1)$.

In the $SO(8)$ example discussed above, we have two Jordan blocks corresponding
to the partition 8 = 5 + 3, $s=2$ and $\bar G_{e,h,f}$ is trivial.

\begin{figure}
    \begin{center}
        \epsfxsize=300pt
        \epsfysize=0pt
        \vspace{-5mm}
        \parbox{\epsfxsize}{\epsffile{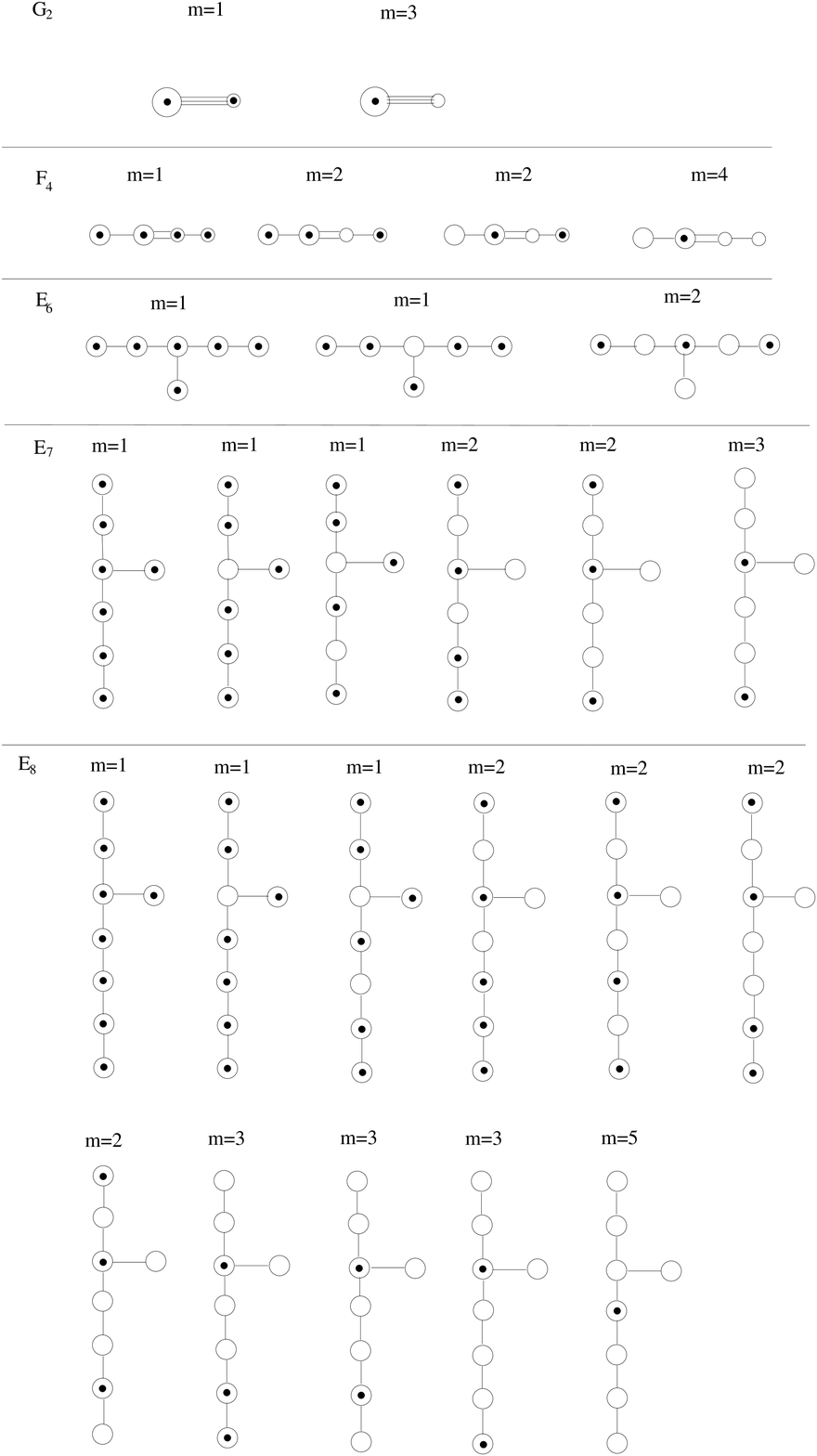}}
        \vspace{5mm}
    \end{center}
\caption{Unbroken discrete subgroups $S_m$ of the exceptional groups for a 
supersymmetric Higgs vacuum state corresponding to a given distinguished
marked Dynkin diagram.}
\label{mexcept}
\end{figure}

A. Alexeevski \cite{Alex} showed that for exceptional Lie algebras, the
groups  $\bar G_{e,h,f}$ are always isomorphic to one of the symmetric 
groups $S_m$ where 
$m = 1,2,3,4$
or 5. We list all the distinguished marked Dynkin diagrams  for exceptional 
Lie algebras and the corresponding  values of $m$ in the table in Fig. 
\ref{mexcept}. Note that in all cases $m=1$ for the universal distingushed 
triples.

\newpage

Let us give the explicit construction of  $\bar G_{e,h,f}$  for $G_2$.
Consider the following elements of the group $G_2$:
  \be
\label{genS3}
a \  &=& \ \exp\{ 2\pi i \beta^\lor /3 \} \nonumber \\
b\ &=& \ \exp \left\{ \frac {\pi i}2 (e_\beta + f_\beta) \right \}
\exp\{ \pi i h/2 \}
  \ee
with $h = 2\alpha^\lor + \beta^\lor$ as in Eq.(\ref{tripG2}). The triple
(\ref{tripG2}) is invariant under the action of $a$ and $b$. For the
element $a$, it follows directly from the standard commutators
$[\beta^\lor, e_\alpha] = -3e_\alpha, \  [\beta^\lor, e_{\alpha + 3\beta}]
= 3e_{\alpha+3\beta}$. The element $b$ realizes the automorphism
\footnote{A side remark is that $b$ is an ``exceptional'' element in the
sense of Ref.\cite{Kac}: the fundamental group of the centralizer
$G_b = [SU(2)]^2/{\bf Z}_2$ is non-trivial (more exactly, the relevant
fact is that the fundamental group  involves a
non-trivial finite factor).} 
  \be
 \label{autob}
\beta^\lor \rightarrow - \beta^\lor,\ \ 
\alpha^\lor \leftrightarrow (\alpha + 3\beta)^\lor,\ \ 
(\alpha+ \beta)^\lor \leftrightarrow (\alpha + 2\beta)^\lor\ ,  \nonumber \\
e_\alpha \leftrightarrow e_{\alpha + 3\beta},\ \ 
e_{\alpha+\beta} \leftrightarrow e_{\alpha + 2\beta},\ \ \ 
f_\alpha \leftrightarrow f_{\alpha + 3\beta},\ \ 
f_{\alpha+\beta} \leftrightarrow f_{\alpha + 2\beta}\ 
  \ee
(the first line in Eq.(\ref{autob}) describes the action of $b$ on the Cartan
subalgebra of ${ g}_2$; it is just the Weyl reflection of the system
of coroots in Fig. 1b with respect to the line orthogonal to  $\beta^\lor$).
Again, the triple (\ref{tripG2}) is invariant.

$a$ and $b$ satisfy the relations 
 \be
\label{relS3}
a^3 = 1,\ \ b^2 = 1, \ \ \ ab = ba^2
 \ee
The first two of them follow immediately from the definition 
$h = 2\alpha^\lor +
\beta^\lor$ and from the property (\ref{unity}), and the third one
is easily verified if leaving aside the common $U(1)$ factor
$ \exp\{ \pi i h/2 \}$ and writing the $SU(2)$ part in the matrix form:
 \be
 a \ \sim  \left(
\begin{array}{cc}  e^{2\pi i/3} &  0 \\
0 & e^{-2\pi i/3}  \end{array} \right),\ \ 
b \ \sim  \left(
\begin{array}{cc}  0 &  i \\ i & 0  \end{array} \right)
 \ee
The elements $a,b$ satisfying the defining relations (\ref{relS3}) generate
the group $S_3$, the centralizer of the distinguished exceptional 
triple (\ref{tripG2}) in $G_2$.

In the physical language, the presence of non-trivial finite 
centralizers $\bar G_{e,h,f}$  means that our Higgs averages, the
solutions to Eq.(\ref{DF0}), in many case break down the gauge symmetry not
completely, but a discrete subgroup of the original gauge group remains
unbroken. Some isolated examples of the systems where this phenomenon takes
place have been discovered before (see e.g. Ref.\cite{Etesi}), but our 
construction with 3 adjoint scalar fields is much more natural and gives a 
rich family of such examples.

\section{Deficit term.}
\setcounter{equation}0

\subsection{Generalities. $SU(2)$ case.}

As we have seen, solving the equation system (\ref{DF0}) is the most direct 
and the most simple way to obtain the answer. But the other ways to 
calculate the index, in particular the traditional way based on the formula
(\ref{princdef}) present also a considerable methodic interest. The principal
contribution has been calculated only for the unitary groups, and we do not
know now how to do it in other cases. Speaking of the deficit term, its 
calculation
is much more simple. In this section, we first describe following 
Refs.\cite{YiSethi,GG} how the deficit term is calculated for the unitary
groups and generalize this calculation to some orthogonal, symplectic, and
exceptional groups. 

The main idea is the following. Let us regularize the theory in infrared
putting it in the large but finite ball with rigid walls 
$(A_I^A)^2 \leq R^2$. Such a regularization breaks down supersymmetry, and the
supertrace ${\rm Tr} \{(-1)^F e^{-\beta H} \} $ acquires a certain 
$\beta$--dependence so that the integral in the second term in 
Eq.(\ref{princdef})
does not vanish.
It turns out that this gives a non--vanishing contribution also in the
limit $R \to \infty$ we are interested in ! 
It is more or less clear (we address the reader to Ref.\cite{YiSethi} for
details) that this is associated with the behavior of the theory at large
$|{\bf A}|$.
\footnote{Another way of reasoning not requiring a rigid ball regularization
and not using Eq.(\ref{princdef}) is the following: The functional integral
for the index in the limit $\beta \to 0$ includes not only the contribution
of possible normalizable supersymmetric vacuum state, but is also 
``contaminated' by the low--lying states of the continuum spectrum. The
deficit term is exactly this continuum--driven contamination whose dynamics
depends on the   large
$ |{\bf A}|$ region.}
 However, as we have already noted,  at large
$ |{\bf A}| \gg g^{-1/3}$ the theory is greatly simplified. Basically, it is
given by the effective hamiltonian in Eq. (\ref{QHeff}).

Strictly speaking, the statement expressed in the last sentence is wrong and
we will see it soon. It is true, however, for, say, $SU(2)$ or $SU(3)$
gauge groups. It makes sense to understand first the spirit of the 
argumentation in the simplest $SU(2)$ example, and then we will easily 
understand how to correct it in more complicated cases.

We will briefly describe now the Born--Oppenheimer approach suggested in 
\cite{Wit},
 developed in details in \cite{chiral,correct,Trieste} and rediscovered
in recent \cite{Halp1,Hoppe1,Konech}. If the gauge group is $SU(2)$ and 
$(A_I^A)^2$ is
large, we can subdivide the physical bosonic variables (there are altogether
24 such variables: $3 \times 9$ modulo 3 gauge degrees of freedom) into two
groups: 9 slow variables $A_I^{3\ {\rm slow}} = c_I$ which describe the 
motion along the vacuum valley and 15 fast variables $A_I^{a\ {\rm fast}}$
with  $A_I^{a\ {\rm fast}} c_I = 0$, \ $a = 1,2$.                          
                   
A meticuluous reader might have been perplexed by the fact that $2 \times 9
- 2 = 16 > 15$, but it is just because not all the variables
 $A_I^{a\ {\rm fast}}$ are physical. In $SU(2)$ case, any solution of the 
valley equation has the form $A^A_I = \eta^Ac_I$. We have used just 2 gauge
parameters to bring it to the form
 $\eta^A = \delta^A_3$. One gauge parameter is left, and it corresponds to the
$U(1)$ rotation of the fast variables  $A_I^{a\ {\rm fast}}$ around the 
third isotopic axis.

Typically,  ${\bf A}^{\rm fast} \ll {\bf A}^{\rm slow}$. Using this, we can
classify the terms in the full hamiltonian by the powers of the formal 
parameter   $|{\bf A}^{\rm fast} |/ |{\bf A}^{\rm slow}|$. The leading term
$H_{0}$  has the form
 \be
 \label{H0}
H_0 \ =\ \left(\delta_{IJ} - \frac{c_I c_J}{{\bf c}^2} \right)
\left(- \frac 12 \frac{\partial^2}{\partial A^a_I \partial A^a_J}
+ \frac {g^2 {\bf c}^2}{2} A^a_I A^a_J \right)
+ \frac{ig}2 c_I \epsilon^{ab} \lambda^a \Gamma_I \lambda^b
 \ee
It is just a supersymmetric oscillator (one should understand that the
hamiltonian (\ref{H0}) acts on fast variables and 
their superpartners, and the slow variables
$c_I$ play the role of parameters)  The ground supersymmetric
state of $H_0$ has zero energy and the gap separating it from excited
states is $\sim g|{\bf c}|$. The total wave function may be written
as
 \be
 \label{Psitot}
\Psi(x^{\rm fast}, x^{\rm slow}) \ =\     \sum_n \chi_n(x^{\rm slow})
\psi_n(x^{\rm fast}) \approx  \nonumber \\
\chi_0(x^{\rm slow})
\psi_0(x^{\rm fast}) + {\rm contribution \ of\  excited \ states} \ ,
 \ee
where $x^{\rm fast}$ stand for $A^{\rm fast}$ and their superpartners,
$x^{\rm slow}$ stand for $A^{\rm slow}$ and their superpartners and
$\psi_n(x^{\rm fast}) \equiv |n\rangle $ is the spectrum of $H_0$.
 
Let us find the explicit expression for  $\psi_0(x^{\rm fast})$ . 
To this end, it is 
convenient to choose $\Gamma_I$ so that
  \be
\label{choice}
 \frac 1{|{\bf c}|} c_I \Gamma_I \ =\ {\rm diag} (\gamma_3, \gamma_3)
\ {\rm with} \ \ \gamma_3 = {\rm diag} (\sigma_3, \sigma_3, \sigma_3, 
\sigma_3) 
  \ee
Then the hamiltonian (\ref{H0}) conserves the ``fast fermion charge''
  \be
 \label{Ffast}
F^{\rm fast} \ =\ \mu_{\tilde \alpha}^a \bar \mu_{\tilde \alpha}^a 
  \ee
with $\mu_{\tilde \alpha}^a$ defined as in Eq.(\ref{mulam}). With this choice,
$\mu_{\tilde \alpha}^a$ are naturally decomposed into 4 sets of variables
$\mu_{1,2}^a \,\ \mu_{3,4}^a \ ,\ \mu_{5,6}^a \ , \ {\rm and}\ 
 \mu_{7,8}^a$\ , each such set 
corresponding to a couple of 4--dimensional
Weyl fermions. Let us introduce the antisymmetric $8 \times 8$ matrix 
of charge conjugation 
$ C \ =\ {\rm diag} (i\sigma_2, i\sigma_2, i\sigma_2, i\sigma_2)$
 (it rises and lowers the indices for each 4D Weyl fermion).
The vacuum wave function reads \cite{Trieste}
  \be
 \label{vacfast}
|0> \ \propto \ |{\bf c}|^4 \exp \left\{ - \frac{g|{\bf c}|}2 A_I^a A_J^a
\left( \delta_{IJ} - \frac{c_I c_J}{|{\bf c}|^2} \right) \right\}
[\mu^a  C  \mu^a  + i \epsilon^{ab} \mu^a C \gamma_3 \mu^b]^4\ ,
  \ee
where the prefactor $|{\bf c}|^4$ makes the normalization integral
$\langle 0|0 \rangle$ ${\bf c}$--independent.  
The effective supercharges  are defined as
 \be
\label{usredQ}  
Q^{\rm eff}_\alpha \ =\ \langle 0| Q_\alpha | 0 \rangle \ ,
  \ee
where $Q_\alpha $ are the full supercharges (\ref{Q}).
To find (\ref{usredQ}), note first that, as the wave function (\ref{vacfast})
has a definite fast fermion charge $F^{\rm fast} = 8$, only the part of 
(\ref{Q}) with  $F^{\rm fast} = 0$ contributes. It has the form
  \be
\label{Qzwisch}
 Q_\alpha \ =\ \frac 1{\sqrt{2}} [(\Gamma_I)_{\alpha \beta} E^3_I 
+ \frac g2 (\Gamma_I \Gamma_J)_{\alpha \beta} \epsilon^{ab} A_I^a A_J^b ] 
\lambda^3_\beta
 \ee
Let us observe now that the second term in Eq.(\ref{Qzwisch}) has zero
average over the  vacuum state (\ref{vacfast}) of $H_0$ due to
$$
\int \left( \prod_{aI}' dA_I^a \right) [\epsilon^{ab} A^a_M A^b_N ]
\exp \left\{ - {g|{\bf c}|} A_I^a A_J^a
\left( \delta_{IJ} - \frac{c_I c_J}{|{\bf c}|^2} \right) \right\} 
\ =\ 0 \ , $$
where the prime in $\prod'$ means that the integral is done only over the
16 fast variables with $(A_I^a)^{\rm fast} c_I = 0$. 

The remaining first term just coincides with the supercharge in Eq.
(\ref{QHeff})  with only one term in the sum. To show that $Q_\alpha^{\rm eff}$
is, indeed, given by this expression, it is necessary, however, to be
convinced that the contribution of the term
when the derivative $\partial/\partial c_I$ acts on the wave function
(\ref{vacfast})  gives zero.
 As far as the bosonic part of Eq.(\ref{vacfast}) is concerned, we have
  \be
\int \left( \prod_{aI}' dA_I^a \right) 
 |{\bf c}|^4
\exp \left\{ - {g|{\bf c}|/2} \ A_I^a A_J^a
\left( \delta_{IJ} - \frac{c_I c_J}{|{\bf c}|^2} \right) \right\} \nonumber \\
\frac \partial {\partial c_I} |{\bf c}|^4
\exp \left\{ - {g|{\bf c}|/2}\  A_I^a A_J^a
\left( \delta_{IJ} - \frac{c_I c_J}{|{\bf c}|^2} \right) \right\} \nonumber \\ 
\ =\  \ \frac \partial {\partial c_I} \ 
\int \left( \prod_{aI}' dA_I^a \right) 
 |{\bf c}|^8
\exp \left\{ - {g|{\bf c}|} A_I^a A_J^a
\left( \delta_{IJ} - \frac{c_I c_J}{|{\bf c}|^2} \right) \right\} \ =\ 0\ .
 \ee

The only thing  which is left to understand is what happens when the derivative
$\partial/\partial c_I$ acts on the fermion part of the wave function
(\ref{vacfast}). The latter depends on $c_I$ because the choice (\ref{choice})
is possible only for some particular value of $c_I$. When $c_I$ is shifted, 
the fermion part of the hamiltonian (\ref{H0}) is modified and so is the 
fermion structure of its vacuum wave function. Let us first consider the 
${\cal N}=1$
theory with the hamiltonian written in Eq.(\ref{HN1}). Slow bosonic variables 
present a 3--vector $c_i$ and we have just a 
couple 
of fast Weyl fermions $\lambda^a_\alpha$, $\alpha = 1,2$. The fermion
structure of the analog of Eq. (\ref{vacfast}) is
  \be
 \label{vacN1}
\psi_0(x^{\rm fast}) \ \propto \ 
\lambda^{a\alpha} \lambda_\alpha^a \ + \ i \epsilon^{ab} \lambda^{a\alpha}
(\sigma_k)_\alpha^\beta \lambda_\beta^b \frac {c_k}{|{\bf c}|} \ ,
  \ee
where $\lambda^{a\alpha} = \epsilon^{\alpha\beta} \lambda_\beta^a
 = (i\sigma_2)_\alpha^\beta \lambda^a_\beta$. Differentiating this over
$c_k$ gives the structure
  \be
  \label{vozb}
\propto \ \epsilon^{ab} \lambda^{a\alpha} (\sigma_j)_\alpha^\beta 
\lambda_\beta^b \left[ \delta_{kj} - \frac {c_k c_j}{|{\bf c}|^2} \right]\ .
  \ee
This structure is orthogonal to (\ref{vacN1}) and gives zero after averaging.

A certain complication of the ${\cal N}=4$ case is due to the fact that 
$\mu_{\tilde \alpha}^a$ do not form a representation of $SO(9)$. For 
a generic $c_I$, the hamiltonian $H_0$ does not conserve the fast fermion
charge (\ref{Ffast}), and the form of the wave function is more compicated than
that in Eq.(\ref{vacN1}). We can make use of the $SO(9)$ invariance, however, 
and
{\it assume} a special form of the shift $\delta c_I$: 
\ $\delta c_4 = \ldots = \delta c_9 = 0$. In 4--dimensional language, that 
means that only the gauge field is shifted and scalar fields are not.
In that case, the vacuum wave function involves the product of 4 factors
like in Eq. (\ref{vacN1}) and its differentiating over $c_i$ produces the
structure (\ref{vozb}) with zero projection on the vacuum state.

Thus, we have shown that
$$ \langle 0 | \partial /\partial c_I | 0 \rangle \ =\  
\partial /\partial c_I $$
 and the effective supercharge (\ref{usredQ}) is given by Eq.(\ref{QHeff}).
We have derived it for the $SU(2)$ theory, but the derivation can be 
easily generalized for other groups. We will comment on that a bit later. 
 
The effective hamiltonian (in the $SU(2)$ case, it is just the 9--dimensional
laplacian) can be obtained as $(1/16)\{Q_\alpha, Q_\alpha\}_+$ or also
determined  with the Born--Oppenheimer procedure (which is a little bit
trickier than for supercharges because one should take into account also
the contribution of excited states  in Eq.(\ref{Psitot}) 
\cite{chiral,correct} ). 
$Q_\alpha^{\rm eff} $ and $H^{\rm eff}$ act on (properly normalized)
$\chi_0(x^{\rm slow})$.

The simple result (\ref{QHeff}) is specific for pure supersymmetric  
Yang--Mills theories and theories with non--chiral matter content. 
For chiral theories (like QED with one left chiral superfield of charge
2 and 8 right chiral superfields of charge 1, or the
$SU(5)$ theory with a left quintet and a right decuplet), the second term
in the analog of (\ref{Qzwisch}) and also the term due to the
action of the derivative $\partial/\partial c_i$ onto
$\psi_0(x^{\rm fast}) $ do not vanish after averaging.  
 A Berry phase
(with singularity at $|{\bf A}| = 0$ )
appears \cite{chiral}. Also, for non--chiral theories, but in the 
next--to--leading Born--Oppenheimer order, the calcutations are 
more involved and the expressions are less trivial, 
the vacuum moduli space acquires a (conformally flat, but not just flat) 
 metric, etc \cite{correct}. All this is irrelevant in our case, however.

As the deficit term is related to the large values of $|{\bf A}|$ and as, for
such large values, our original problem is equivalent to (\ref{QHeff}),
the deficit term of the original hamiltonian should be equal to the deficit
term of $H^{\rm eff}$. 
$H^{\rm eff}$ describes free motion and its spectrum 
consists of delocalized plane waves. Obviously, the total index 
(\ref{princdef}) of $H^{\rm eff}$ is zero. And that means that the deficit 
term for 
$H^{\rm eff}$ (and hence the deficit term for original hamiltonian 
(\ref{H}) should coincide with the principal contribution in $H^{\rm eff}$.

Naively, the latter seems to be zero. Indeed, the classical hamiltonian
$H^{\rm eff} = P_I P_I/2  $ does not depend on fermion variables, and the 
fermion integrals in the analog of Eq.(\ref{intind}) should give zero.
This is not true, however, and the reason is that not all eigenstates
of $H^{\rm eff}$ are physical and contribute in the supertrace
${\rm Tr}\{(-1)^{-1} e^{-\beta H} \}$. The matter is that the requirement of
the gauge symmetry of the wave functions imposes the requirement of
{\it Weyl invariance} on the eigenstates of the effective hamiltonian
\cite{Wit}. In the $SU(2)$ case, this  Weyl invariance corresponds
just to the simultaneous sign reversal for $c_I$ and $\lambda_\alpha$. The 
wave functions should not change under such a transformation. 

To be more precise, wave functions depend on $c_I$ and holomorphic 
 variables $\mu_{\tilde \alpha}$ defined like in Eq.(\ref{mulam}).
 In 4--dimensional language, 8 complex
variables can be interpreted as 4 Weyl $4D$ spinors $\psi_{\alpha f}$
with $\alpha = 1,2$ and $f = 0,1,2,3$ ($\psi_{\alpha 0}$ is the abelianized
version of gluino and  $\psi_{\alpha(1,2,3)}$ are related to the fermion
components of chiral matter multiplets). Thus, we impose the requirement
 \be
 \label{Weylsu2}
\Psi(-c_I, -\mu_{\tilde \alpha}) \ =\ \Psi(c_I, \mu_{\tilde \alpha})
  \ee
It is not difficult to implement the discrete symmetry (\ref{Weylsu2})
in the path integral language. The supertrace projected on the states
invariant under the action of the Weyl symmetry  (\ref{Weylsu2}) can be 
presented as the integral
  \be
 \label{IndK2}
 I_W \ =\ \frac 12 \int \prod_I {dc_I}
\prod_{{\tilde \alpha} } d\bar \mu_{\tilde \alpha} d\mu_{\tilde \alpha} \exp
\left\{ -  \bar \mu_{\tilde \alpha} \mu_{\tilde \alpha} \right  \} \nonumber \\ 
\left[ 
{\cal K} ( c_I,\bar \mu_{\tilde \alpha}; \  c_I, \mu_{\tilde \alpha};\ \beta) +
{\cal K} ( -c_I, -\bar \mu_{\tilde \alpha}; \  c_I, 
\mu_{\tilde \alpha};\ \beta) \right]\ ,
 \ee
where ${\cal K} (\cdots)$ is the kernel of the evolution operator. In our 
case (free motion !), this kernel has the simple form
  \be
 \label{kernel}
{\cal K} ( c_I', \bar \mu_{\tilde \alpha}';  c_I,\ \mu_{\tilde \alpha};\ \beta)
\ =\ \frac{1}{\sqrt{2\pi \beta}}\exp \left\{ \bar \mu_{\tilde \alpha}' \mu_{\tilde \alpha}
 -\frac{ (c_I' - c_I)^2}{2\beta}  \right\}
  \ee
Substituting it in Eq.(\ref{IndK2}), we find that: {\it (i)}
the first term $\equiv$ supertrace for the system with no constraints imposed
= 0;  {\it (ii)} The only contribution comes from the second term, and it
is non--zero:
 \be
 \label{detsu2}
I_W^d = \frac 12 \frac{2^8}{2^9} = \frac 14
  \ee
The factor $2^8$ in the numerator comes from the fermion integrals and the
factor $2^9$ in the denominator --- from the bosonic integrals.

Note that the result (\ref{detsu2}) holds universally for ${\cal N}=4$ theory
and also for ${\cal N}=2$ and ${\cal N}=1$ theories. Indeed, for ${\cal N}=1$, we have just two
fermionic variables and 3 bosonic variables, for ${\cal N}=2$, we have 4 
fermionic variables and 5 bosonic variables. We obtain
 $$I_W^d = \frac 12 \frac{2^8}{2^9} = \frac 12 \frac{2^4}{2^5} 
= \frac 12 \frac{2^2}{2^3}\ = \frac 14 $$
for any ${\cal N}$.

\subsection{Proper deficit term.   }

What happens for other groups ? Let us start with applying the same logics 
and calculate
the deficit term in the original nonabelian theory as  the principal 
contribution in the abelian effective theory (\ref{QHeff}) where only the Weyl
 invariant states
 \be
 \label{Weylsym}
\Psi(w c_I^s, w \mu_{\tilde \alpha}^s) \ =\ \Psi(c_I^s, \mu_{\tilde \alpha}^s)
  \ee
are taken into account. 
Let us call this contribution the ``proper deficit term'' (we will
see by the end of this section that it is not the whole story and, 
generically, some other contributions appear.) 

The generalization of Eq.(\ref{IndK2}) can be easily
written:
  \be
 \label{IndK}
 I_W \ =\ \frac 1{\# W} \int \prod_{Is} {dc_I^s}
\prod_{{\tilde \alpha} s} d\bar \mu_{\tilde \alpha}^s d\mu_{\tilde \alpha}^s 
\exp
\left\{ -  \bar \mu_{\tilde \alpha}^s \mu_{\tilde \alpha}^s \right  \} 
\sum_{w \in W}
{\cal K} ( w c_I^s, w \bar  \mu_{\tilde \alpha}^s; \  c_I^s, 
\mu_{\tilde \alpha}^s;\ \beta) \ .
 \ee
Substituting here the corresponding generalization of (\ref{kernel}) and 
doing the integrals, we obtain
 \be
\label{IndWeyl}
 I_W \ =\ \frac 1{\# W} \sum_{w \in W} ' \frac 1{\det(1-w)}\ ,
  \ee
where $\Sigma'$ means that the sum is done over all elements of $W$ with 
${\rm det} (1-w) \neq 0$.
Again, this is true for all ${\cal N}$. 

It is instructive to compare the result (\ref{IndWeyl}) with the formula
 \be
\label{Indtor3}
 I_W \ =\ \frac 1{\# W} \sum_{w \in W} [\det(1-w)]^2
  \ee
which counts the number of vacuum states within the sector of constant gauge
potentials for the ${\cal N}=1$ SYM theory defined on small 3--torus.
\footnote{Note in passing that  for higher orthogonal and for 
exceptional groups, 
there are also other vacuum states associated with certain non-trivial triples
of commuting group elements \cite{Wittrip,Kac}.}
We have here the same set of variables $c_i^s,\ \psi_\alpha^s$ ($i=1,2,3$), 
but the motion in the space of $c_i^s$ is finite, and the spectrum is 
discrete. For {\it vacuum} states, the  wave functions  do not depend at 
all on
$c_i^s$, but only on $\psi_\alpha^s$. The index can be presented in the 
integral form (\ref{IndK}) where the evolution operator depends now only
on fermion variables (and there is no integral over $\prod dc_i^s$). Thus,
instead of $\det^2(1-w)/\det^3(1-w)$ we have just $\det^2(1-w)$ and 
Eq.(\ref{Indtor3}). 
  \footnote{There are at least two other ways to derive (\ref{Indtor3}). First,
one can duly take into account the dependence of ${\cal K}$ on bosonic 
dynamical variables, but implementing besides (\ref{Weylsym}) also the
periodicity conditions
 $$
\Psi( c_I^s + L_i^s) \ =\ \Psi(c_I^s)\ 
  $$
where $L_i^s$ is a set of 3 $s$--dimensional vectors leaving invariant the Weyl
lattice. In other words, for the theory defined on the torus, global structure
of our gauge group becomes important, and the Weyl symmetry involves not 
only rotations, but also translations [for $SU(2)$ we have, besides the
symmetry $c_i, \lambda_\alpha \to -c_i, -\lambda_\alpha$, also the symmetry
$c_i \to c_i + 2\pi n_i /(gL)$ with integer $n_i$]. One can be convinced
that the effects brought about by the Weyl rotations $c_i^s \to w c_i^s$
and the Weyl translations $ c_i^s \to  c_i^s + L_i^s $ exactly cancel each
other, and the answer (\ref{Indtor3}) is reproduced.

Formula (\ref{Indtor3})  is also well  known
to mathematicians. It gives the number of invariants of $W$ in the Grassman
algebra over ${\mathfrak h} \times {\mathfrak h}$ which is equal to $r+1$
due to \cite{Sol} (see \cite{Kac}). Incidentally, Eq.(\ref{Indtor3}) with
2 replaced by 1 is equal to 1 since $W$ has no non--trivial invariants
in the Grassman algebra over ${\mathfrak h}$ \cite{Sol}.  } 

Similar formulae counting the number of vacuum states in 6D and 10D SYM 
theories defined on 5--dimensional and 9--dimensional spatial tori, 
respectively, can be written:
 \be
\label{Indtor59}
D=6:\ \ \ \ \  I_W \ =\ \frac 1{\# W} \sum_{w \in W} [\det(1-w)]^4
\nonumber \\
D=10:\ \ \ \ \  I_W \ =\ \frac 1{\# W} \sum_{w \in W} [\det(1-w)]^8
  \ee

The problem of counting the states on 3--torus is especially simple. One can 
show that the  sum in the R.H.S. of Eq.(\ref{Indtor3}) is equal to $r+1$
for any group \cite{Kac} which justifies the original Witten's conjecture
\cite{Wit}. There is no such simple universal formula  for the sum 
(\ref{IndWeyl}), however. One should calculate it case by case.

Consider first the higher unitary groups \cite{GG}. The Cartan subalgebra
of $su(n)$ is realized by the matrices diag$(a_1,\ldots,a_n)$, $a_1 + \ldots +
a_n = 0$. The Weyl group is the group $S_n$ of permutations of $a_n$.
Only the elements $w \in W$ with $\det (1-w) \neq 0$ contribute in the
sum (\ref{IndWeyl}). These are the so called {\it Coxeter elements} 
corresponding to the cyclic permutations of $\{a_n\}$. There are
$(n-1)!$ such elements (while   $\# W = n!$). All they are 
equivalent by conjugation and have one and the same  $\det (1-w)$. Take the
element
  \be
\label{Coxet}
  w_*\ =\ \left( \begin{array}{ccccc}
0&1&0&\ldots&0 \\ 0&0&1&\ldots&0 \\ \ldots&\ldots&\ldots&\ldots&\ldots \\
0&\ldots&\ldots&\ldots&1 \\ -1&-1&\ldots&\ldots&-1 \end{array} \right)
  \ee
(this is just the permutation $a_1 \to a_2 \to \ldots \to -a_1 - \ldots 
-a_{n-1} \to a_1$ written as a matrix in the basis $\{a_1,\ldots,a_{n-1}\}$.
 One can show that  $\det (1-w_*) = n$. Therefore, the proper deficit
term for the $SU(n)$ gauge group is
 \be
\label{inhdsun}
 I_W^{\rm prop.\ def.} [SU(n)] \ =\ \frac 1{n^2}
  \ee
As a simple exercise, one can calculate the sums in 
Eqs.(\ref{Indtor3},\ \ref{Indtor59}) and derive
\footnote{It is remarkable that the same result for the D=10 theory is 
obtained with the t'Hooft twisted boundary conditions \cite{Gomez}. }
 \be
 \label{torisun}
D=4:\ \ \ \ \ \ I_W = n \nonumber \\
D=6:\ \ \ \ \ \ I_W = n^3 \nonumber \\
D=10:\ \ \ \ \ \ I_W = n^7 
  \ee

Let us go over now to non--unitary groups. Consider first the groups
$SO(2r+1)$ and $Sp(2r)$ [the root systems for these groups are dual to each 
other, their Weyl
groups coincide, and the proper deficit term (\ref{IndWeyl}) is the same].
It is easier to think in symplectic language. The elements of Cartan subalgebra
of $sp(2r)$ can be represented as the diagonal matrices diag $(a_1,\ldots,a_r,
-a_1,\ldots,-a_r)$. The Weyl group is the product of the group $S_r$
of permutations of $a_i$ and $r$ $Z_2$ factors corresponding to the
reflections $a_i \to -a_i$. The essential complication compared to the
unitary case is that many different conjugacy classes contribute in the
sum (\ref{IndWeyl}). In the simplest case of $Sp(4) = SO(5)$, $\# W = 8$,
and 3 elements of two different conjugacy classes contribute:
 \be
\label{SO5}
w \ =\ \left( \begin{array}{cc} -1&0 \\ 0&-1 \end{array} \right), \ \ 
w \ =\ \left( \begin{array}{cc} 0&1 \\ -1&0 \end{array} \right), \ \ 
w \ =\ \left( \begin{array}{cc} 0&-1 \\ 1&0 \end{array} \right) \ . 
 \ee
That gives 
  \be
\label{IndSO5}
  I_W^{\rm prop.\ def.} [SO(5)] \ =\ \frac 18 \left( \frac 14 + \frac 12 
+ \frac 12 \right) \ =\ \frac 5{32}
 \ee
We have also made explicit calculations for $Sp(6)$, $SO(7)$, and
$G_2$. The results are given in Table 2 below.

\subsection{Total deficit vs. proper deficit.}

Eq.(\ref{inhdsun}) is the correct result for the deficit term for the unitary
groups when $n$ is prime. But if $n$ is not prime, Eq.(\ref{inhdsun}) is only
one of the contributions. The total deficit term is given by the sum 
(\ref{defres})
over the divisors of $n$. In Ref.\cite{GG}, the appearance of these extra terms
was explained in terms of D--particles and D--instantons. We present here a
conventional (or, better to say, a conservative) explanation.

Some problem appears already on the level of $SU(3)$. The vacuum valley is 
labelled by  two  9--vectors $a_I$ and $b_I$ so that 
   \be
  \label{valsu3}
 (A^A_I)^{\rm slow} t^A \ =\ {\rm diag} (a_I, b_I - a_I, -b_I)
   \ee
Let us find now the $SU(3)$ version of (\ref{H0}) . To this 
end, we substitute ${\bf A} = {\bf A}^{\rm slow} + {\bf A}^{\rm fast}$
in the hamiltonian (\ref{H}) and pick up the leading terms in
$|{\bf A}^{\rm fast}|/| {\bf A}^{\rm slow}|$. The potential part of $H_0$ 
reads
\cite{Trieste}
 \be
\label{V03}
V_0 \ =\ \frac{g^2(2{\bf a} - {\bf b})^2}2 A^{a = 1,2}_I  A^{a = 1,2}_J
\left(\delta_{IJ} - \frac{(2a-b)_I (2a-b)_J}{{\bf (2a-b)}^2} \right)\ +
\nonumber \\
\frac{g^2({\bf a} + {\bf b})^2}2 A^{a = 4,5}_I  A^{a = 4,5}_J
\left(\delta_{IJ} - \frac{(a+b)_I (a+b)_J}{{\bf (a+b)}^2} \right) \ +
\nonumber \\
\frac{g^2(2{\bf b} - {\bf a})^2}2 A^{a = 6,7}_I  A^{a = 6,7}_J
\left(\delta_{IJ} - \frac{(2b-a)_I (2b-a)_J}{{\bf (2b-a)}^2} \right)
  \ee
Thus, there are 48 fast variables divided naturally into three groups:
$A_I^{1,2}$ satisfying the condition $(2{\bf a} - {\bf b}){\bf A}^{1,2} = 0$,
$A_I^{4,5}$ satisfying the condition $({\bf a} + {\bf b}){\bf A}^{1,2} = 0$,
and 
$A_I^{6,7}$ satisfying the condition $(2{\bf b} - {\bf a}){\bf A}^{1,2} = 0$.
(six variables with nonzero projection $(2{\bf a} - {\bf b}){\bf A}^{1,2}$,
etc. are the gauge degrees of freedom; two other gauge degrees of freedom
are ``hidden'' in 48 variables: we should require that
 eigenstates of $H_0$ be annihilated by the Gauss constraints
$G^3$ and $G^8$ ).

Again we have a  supersymmetric oscillator. Or rather combination
of several oscillators with frequencies $g|2{\bf a} - {\bf b}|$,
$g|{\bf a} + {\bf b}|$ and $g|2{\bf b} - {\bf a}|$.
The standard Born--Oppenheimer 
philosophy  of Refs.\cite{Wit,Trieste} works when these frequencies are much 
larger that the 
characteristic energy scale $g^{2/3}$, i.e. when 
$$g|2{\bf a} - {\bf b}|^3 \gg 1, \ \ \ g|{\bf a} + {\bf b}|^3 \gg 1,\ \ \ 
g|2{\bf b} - {\bf a}|^3 \gg 1 $$
In other words, the eigenvalues ${\bf a}$, ${\bf b} - {\bf a}$, and
$- {\bf b}$ should not be very small by absolute value, and also they should
not be too close to each other. From mathematical viewpoint, the condition
for the eigenvalues in Eq.(\ref{valsu3}) are different means that the 
centralizer of the generic element (\ref{valsu3}) in (9 copies of) $su(3)$ 
coincides with the (9 copies of) the
 Cartan subalgebra of $su(3)$.

When, say, $2{\bf a} - {\bf  
b} \sim 0$, the fields $A^{1,2}_I$ become massless.
At the point $2{\bf a} = {\bf  b}$, they form together with the fields
$A^3_I$ the $SU(2)$ gauge multiplet. A very important point is that though the
{\it standard} Born--Oppenheimer approach breaks down here, we still {\it can} 
treat the system in   Born--Oppenheimer spirit, only the classification
of the dynamic variables into fast and slow categories is modified. We have
now 9 slow variables 
  \be
  \label{subval3}
 (A^A_I)^{\rm slow} t^A \ =\ {\rm diag} (a_I, a_I , -2a_I)
   \ee
We still have ``abelian'' fast variables $A^{4,5}_I$ and $A^{6,7}_I$.
There are 32 such variables (${\bf A}^{4,5}{\bf a} = {\bf A}^{6,7}{\bf a} = 
0$ !). They 
involve 31 physical variables and a gauge degree of freedom associated with
the rotation around the color axis 8. Besides,  we have 27 variables 
 $A^{1,2,3}_I$
which can be called ``semi-fast'' (we will see very soon why). These 27
variables involve 24 physical semi-fast variables and 3 gauge degrees of 
freedom.
\footnote{All together: $9^{\rm slow} + (31)^{\rm physical\ fast} 
+ (24)^{\rm physical \ semi-fast} + 8^{\rm gauge} = 72$ as it should be.}
 The total wave function can be written as [cf. Eq.(\ref{Psitot})] 
\be
 \label{Psitot3}
\Psi({\bf a}, {\bf A}^{1,2,3},  {\bf A}^{4-7} ) \ =\     
\chi_0({\bf a})
\psi_0^{\rm non-ab}({\bf A}^{1,2,3})  \psi_0^{\rm osc}({\bf A}^{4,5})  
 \psi_0^{\rm osc}({\bf A}^{6,7}) \nonumber \\
+ {\rm contribution \ of\  excited \ states} \ ,
 \ee
(we did not display here explicitly the dependence on the fermion 
superpartners, but one should remember that they are also present). 
$ \psi_0^{\rm osc}({\bf A}^{4,5})$ and $ \psi_0^{\rm osc}({\bf A}^{6,7})$
are the familiar oscillator wave functions (\ref{vacfast}) with 
${\bf c} = 3{\bf a}$. The characteristic scale
 of
$|{\bf A}^{4,5}|, \ |{\bf A}^{6,7}|$ is $\sim 1/\sqrt{g|{\bf a}|}$. And
$\psi_0^{\rm non-ab}({\bf A}^{1,2,3})$ is the wave function of the normalized
vacuum state of the $SU(2)$ theory the existence of which we have 
established before !
The characteristic scale of $|{\bf A}^{1,2,3}|$ is $\sim g^{-1/3}$ which
is still much smaller than $|{\bf a}|$ if $g|{\bf a}|^3 \gg 1$. The hierarchy
 \be
 \label{hier}
\frac 1{\sqrt{g|{\bf a}|}} \ \ll\ \frac 1{ g^{1/3}} \ \ll |{\bf a}|
 \ee
explains why we called the variables ${\bf A}^{1,2,3}$ ``semi-fast''. 
But as far as 
the Born--Oppenheimer method is concerned, there is no distinction between
``fast'' and ``semi-fast'' variables. Once the condition $g|{\bf a}|^3 \gg 1$
is satisfied, we are allowed to integrate out both   ${\bf A}^{4-7}$ and
${\bf A}^{1,2,3}$ and write down the effective theory for  ${\bf A}^{8} \propto
{\bf a}$. Again, this theory is just (\ref{QHeff}) with only one term in the 
sum and describes free motion in 9--dimensional space.

A very important distinction compared with the $SU(2)$ case is, however, that
we {\it should} not impose now the invariance requirement 
with respect to any kind of discrete symmetry for the effective wave functions
$\chi_0(a_I, \lambda_\alpha)$. Indeed, non--trivial elements of the Weyl group
do not leave the subspace (\ref{subval3}) invariant. Once we have fixed the 
gauge as in Eq.(\ref{subval3}), there is no more freedom. As a result, the
valley (\ref{subval3}) gives zero contribution to the index, and the only
non-zero one comes from the generic valley (\ref{valsu3}) with oscillator
wave functions in the fast sector.

The first example where this effect of extra subvalleys with non-abelian fast
sector provides a contribution in the index is the $SU(4)$ theory. Consider
the subspace
 \be
  \label{subval4}
 (A^A_I)^{\rm slow} t^A \ =\ {\rm diag} ({\bf a},{\bf a}  , -{\bf a}, -{\bf a})
   \ee
It presents 9 copies of a certain subalgebra ${\mathfrak h}_a = 
 {\rm diag} (a,a,-a,-a) $ of the Cartan subalgebra of $su(4)$.  The 
centralizer of ${\mathfrak h}_a$ in $su(4)$ is  $su(2) \times su(2) \times
u(1)$ and involves non--abelian factors which support localized vacuum states.
On the other hand, there exist now a non--trivial subgroup $W_a$ of the
original Weyl group $W = S_4$ leaving the 
subalgebra  ${\mathfrak h}_a$ invariant
and acting on its elements faithfully. This is just $Z_2$ a non-trivial
element $w_*$ of which corresponds to the sign reflection $a \to -a$. Thereby,
we {\it have} to impose now the symmetry requirement like in 
Eq.(\ref{Weylsym}).  In the
linear basis in ${\mathfrak h}_a$, $w_* = -1$ and hence $\det(1-w_*) = 2 \neq 
0$.
As a result, the effective theory on the subvalley (\ref{subval4}) has a 
non--zero index (=1/4), and the total deficit term is
  \be
 \label{totdef4}
I_W^{\rm tot \ def}[SU(4)] \ =\ \left( \frac 1{16} \right)_{\rm generic
\ valley}
\ +\ \left( \frac 1{4} \right)_{\rm subvalley\ (\ref{subval4})}
  \ee
For $SU(4)$, there is no other subvalley giving a non--trivial contribution
in the index.

Let us have an arbitrary group $G$, its Lie algebra ${\mathfrak g}$ and the
Cartan subalgebra ${\mathfrak h}$. Let us formulate the conditions on the
subalgebra ${\mathfrak h}_a \subset {\mathfrak h}$ for the subvalley
associated with ${\mathfrak h}_a$ provide a non--zero contribution in the
index:

 {\it (i)}. The centralizer of ${\mathfrak h}_a$ in ${\mathfrak g}$ should
involve a semi-simple factor .
 
 {\it (ii)}. Consider a subgroup $W_a$ of the Weyl group $W$ leaving
invariant  ${\mathfrak h}_a$ . There should be at least one element $w \in
W_a$ such that $\det(1-w) \neq 0$ ($w$ is understood as a matrix in the basis
on ${\mathfrak h}_a$, not ${\mathfrak h}$). 
 
 {\it (iii)}  The center of the centralizer of 
${\mathfrak h}_a$ in ${\mathfrak g}$ should coincide with  ${\mathfrak h}_a$.

{\it (i)} is necessary for the presence of a localized vacuum state in the fast
sector.
\footnote{We want to emphasize here that nonzero contributions from
subvalleys like in Eq.(\ref{subval4}) to the deficit term are specific
for the ${\cal N} =4$ theory. In the ${\cal N} = 1$ and ${\cal N} =2$ cases,
 such localized vacuum states do not appear, and the total deficit term
coincides with the proper one.}
 {\it (ii)} is required for the effective theory to have a non--zero
index. The condition {\it (iii)} guarantees that our Born--Oppenheimer
separation of the variables is justified and that all non-valley field 
variables are indeed fast.
\footnote{ An example of
${\mathfrak h}_a$ which fits the conditions {\it (i)}, {\it (ii)}, but does
not fit the condition {\it (iii)} is ${\mathfrak h}_a =
 {\rm diag} (0,0,a,-a) \in {\mathfrak h}[su(4)] $. Here the variables
$ {\rm diag} ({\bf b},{\bf b},-{\bf b} ,-{\bf b}) $ are exactly as slow
as $ {\rm diag} (0,0,{\bf a},-{\bf a})$. That means that the system does
not want to stay on the valley   $ {\rm diag} (0,0,{\bf a},-{\bf a})$, but
smears out along a larger valley  $ {\rm diag} ({\bf b},{\bf b},{\bf a}-
{\bf b},-{\bf a}-{\bf b})$. This latter valley does not fit, however, the
condition {\it (ii)}.}

{\bf Theorem 7}. Let   ${\mathfrak g} = su(n)$. The only subalgebras 
${\mathfrak h}_a$ of its Cartan subalgebra satisfying the conditions
{\it (i)} -- {\it (iii)} are the Cartan subalgebras of $su(m)$, $m|n$.
The corresponding subvalley gives the contribution $1/m^2$ to the deficit
term in the index.

{\bf Proof}.  
The conditions {\it (i)} and {\it (iii)} imply that 
${\mathfrak h}_a$ presents a subalgebra of ${\mathfrak h}$ commuting with
a certain non--trivial set of root vectors. For $su(n)$ an element of
${\mathfrak h}$ is represented by a traceless diagonal matrix. A subalgebra 
${\mathfrak h}_a$ consists  of such matrices for which some of the elements
are equal, i.e. matrices of the form
$(a,\ldots,a,b,\ldots,b,c,\ldots,c,\ldots)$ where $a$ is repeated 
$k_1$ times, $b$ 
repeated $k_2$ times, etc. To fulfill the condition {\it (ii)}, a permutation
which maps this  ${\mathfrak h}_a$ in itself and has no fixed vectors should
exist. That implies $k_1=k_2=\ldots = k$. The relevant subgroup $W_a$  
is $S_m$ with $k = n/m$: it involves permutations of the set $(a,b,c,\ldots)$.
The corresponding contribution to the index is $1/m^2$.
  Thereby, the result (\ref{defres}) for the total deficit term for the unitary
groups is proven.

\vspace{.3cm}

The subalgebras ${\mathfrak h}_a$ can be found and their contribution to the
deficit term can be calculated also for non--unitary groups.  
As was just noted, the property equivalent to {\it (i)}, {\it (iii)} is that 
there exists a subset of
roots ${\underline{a}} \subset \Delta$  such that 
${\mathfrak h}_a = \{h \in {\mathfrak h} |\alpha(h) = 0 \ {\rm for\ all}
\ \alpha \in {\underline{a}} \}$. It is easy to see that, up to
$W$--conjugacy (conjugacy by an element of the Weyl group), we may choose
 $\underline{a}$ to be a subset of simple roots. 
Given a subset
${\underline{a}}$ of the set of simple roots, let $W_a = \{w \in W|\  
w({\mathfrak h}_a)
\subset {\mathfrak h}_a \}$ and let 
  \be
\label{IndWeyla}
 \left[ I_W^{\rm def} \right]_a \ =\ \frac 1{\# W_a} \sum_{w \in W_a} ' 
\frac 1{\det(1-w)}\ ,
  \ee
Then the total deficit term is given by the sum
   \be
\label{sumWeyla}
 I_W^{\rm tot\ def} =\ \sum_{a \ {\rm mod}\ W}
\left[ I_W^{\rm def} \right]_a 
  \ee
Here the summation is taken over all subsets ${\underline{a}}$ of the set of simple 
roots
modulo $W$--equivalence.  We will calculate the sum (\ref{sumWeyla})
 for the (non--unitary) groups of the 
second and of the third rank.

\begin{figure}
    \begin{center}
        \epsfxsize=250pt
        \epsfysize=0pt
        \vspace{-5mm}
        \parbox{\epsfxsize}{\epsffile{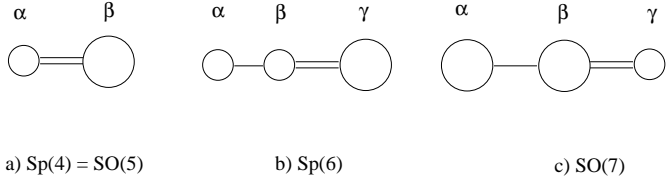}}
        \vspace{5mm}
    \end{center}
\caption{ Dynkin diagrams for some groups .}
\label{Sp46}
\end{figure}

$sp(4)$.  A generic element of 
${\mathfrak h}$ can be presented as a diagonal $su(4)$ matrix \\
diag$(a,b,\ -b,-a)$.  The Dynkin diagram is depicted in Fig.\ref{Sp46}a.
 The corresponding
coroots are $\alpha^\lor = {\rm diag}(1,-1,\ 1,-1)$ and 
$\beta^\lor = {\rm diag}(0,1,\ -1,0)$
There are two different
nontrivial subalgebras  ${\mathfrak h}_a = {\rm diag}(a,a,\ -a,-a)$ 
and  ${\mathfrak h}_a = {\rm diag}(a,0,\ 0,-a)$ 
corresponding to the choice
${\underline{a}} = \alpha$ and ${\underline{a}} = \beta$, respectively. In both cases, 
$W_a = {\bf Z}_2$
(the non--trivial element of $W_a$ being $w:\{a \to -a\}$ ) 
and the contribution (\ref{IndWeyla}) to the index is equal to $1/4$. Adding
it with the proper deficit term from the second line of Table 2
, we obtain the result $21/32$ quoted in the third line.

$G_2$.   Again, we have a long and a short simple root, and
  two different
nontrivial subalgebras  ${\mathfrak h}_a$ . Again, in both cases, 
$W_a = {\bf Z}_2$
giving the contribution  $1/4$. The total deficit term 
is $1/4 + 1/4 + 35/144 = 107/144$.

 $sp(6)$.  A generic element of 
${\mathfrak h}$ can be presented as a diagonal $su(6)$ matrix \\
diag$(a,b,c,\ -c,-b,-a)$.  The Dynkin diagram is depicted in Fig.\ref{Sp46}b.
 The 
coroots are listed in Eq.(\ref{corootSp}).
There are five different
nontrivial subalgebras  ${\mathfrak h}_a$ listed in Table 1
together with the semi--simple parts $G_a$ of the centralizers of
 ${\mathfrak h}_a$  in $G$. \footnote{Note that, in the considered cases, 
the unbroken
gauge group $G_a$ supports only one localized vacuum state. When $G$ is larger,
the number $\#_{\rm vac}[G_a]$ can turn out to be be greater than 1 in which 
case the contribution of the corresponding 
subvalley to the deficit term should be multiplied by $\#_{\rm vac}[G_a]$.} 
The total deficit term is $139/128$.

\begin{table}
\label{TabSp6}

\begin{tabular}{||c|c|c|c|c||} 
set ${\underline{a}}$ &  ${\mathfrak h}_a$  & $W_a$ & $G_a$ & $[ I_W^{\rm def} ]_a$ 
 \\ \hline
$\alpha$ & diag$(a,a,b,\ -b,-a,-a)$& ${\bf Z}_2 \times {\bf Z}_2 $ & 
$SU(2)$ &
1/16 \\ \hline
$\gamma$ & diag$(a,b,0,\ 0,-b,-a)$& $W_{sp(4)} $ & $SU(2)$ &
5/32 \\ \hline
$ \{\alpha,  \beta\}$  & diag$(a,a,a,\ -a,-a,-a)$& ${\bf Z}_2  $ & 
$SU(3)$ & 1/4 \\ \hline
 $\{\alpha,  \gamma\}$ & diag$(a,a,0,\ 0,-a,-a)$& ${\bf Z}_2 $ & 
$[SU(2)]^2$ &
1/4 \\ \hline
  $\{\beta, \gamma\}$ & diag$(a,0,0,\ 0,0,-a)$& ${\bf Z}_2  $ & 
$Sp(4)$ &
1/4 \\ \hline

\end{tabular}

\caption{Subalgebras ${\mathfrak h}_a$ and their contributions in the
deficit term for the $sp(6)$ algebra.}

\end{table}

 $so(7)$.  A generic element of ${\mathfrak h}$ 
has the form $(a,b,c) \ \equiv\ a T_{12} + b T_{34} + c T_{56}$ where 
$T_{ij}$ is the generator of rotation in the $ij$ plane.
 The Dynkin diagram is depicted in Fig.\ref{Sp46}c.
 The corresponding
coroots are $\alpha^\lor = (1,-1,0)$,
 $\beta^\lor = (0,1,-1)$, and  $\gamma^\lor = (0,0,2)$.
Again, there are five different
nontrivial subalgebras  ${\mathfrak h}_a$ associated with the sets
 $\{\alpha\},\ \{\gamma\},\ \{\alpha,  \beta\},\ \{\alpha,  \gamma\}$, and 
$\{ \beta,\gamma\}$. Their contributions to the deficit term are exactly the
same as in the $sp(6)$ case (which is not surprising as the sets of roots
of the algebras $sp(6)$ and $so(7)$ are dual to each other) and 
the total deficit term is  $139/128$.

Adding the total deficit term and the number of normalized vacua determined
earlier with the mass deformation method, we obtain the predictions for the
principal contribution to the index. It would be interesting to confirm them
calculating directly the corresponding integrals. If not analytically
(that must be a difficult task), then first numerically as it was earlier
done for the unitary groups \cite{Stau}. 
 
\begin{table}
\label{inhind}

\begin{tabular}{||c|c|c|c|c||} 
\hline \\

\begin{tabular}{c} gauge \\group \end{tabular} & $\begin{array}{c}
 Sp(4) \\ SO(5) \end{array}$ &  $G_2$ & $ Sp(6)$ & $ SO(7)$
 \\ \hline
\begin{tabular}{c} proper \\ def. term \end{tabular} & $   5/{32}$ &  
$   {35}/{144}$&  $   {15}/{128}$ &  $  {15}/{128}$
\\ \hline
\begin{tabular}{c} total \\ def. term \end{tabular}  & $   {21}/{32}$  &
$   {107}/{144}$&  $   {139}/{128}$ &  $  {139}/{128} $
\\ \hline
$\#_{vac}$ & 1 & 2 & 2 & 1 \\ \hline
\begin{tabular}{c} prediction  \\ for $I_W^{\rm princ.}$ \end{tabular} &
$ {53}/{32} $ &
$  {395}/{144}$ & $  {395}/{128}  $ & $  {267}/{128}$ 
\\ \hline
\end{tabular}

\caption{Deficit term for some groups.}

\end{table}

\section{Asymptotic wave function.}
\setcounter{equation}0

Besides two methods discussed above, the mass deformation method and the
functional integral method, there is also a third way to detect the presence
of the localized supersymmetric vacuum state in the hamiltonian (\ref{H}). 
One can study the solutions of the equation $Q_\alpha |{\rm vac} \rangle = 0$
in the asymptotic region $g|{\bf A}|^3 \gg 1$ where the dynamics is described
by the effective theory (\ref{QHeff}). 

To understand better the philosophy of this method, consider at first a toy
model. Suppose we want to find the localized zero-energy $s$--wave solution
of the Schr\"odinger equation
  \be
 \label{Schr}
\left[ - \frac 12 \Delta + V(r) \right] \psi(r) \ = \nonumber \\
 - \frac 12 \left[ \frac 1{r^{(d-1)/2}} \frac {\partial^2}{\partial r^2} 
\left( r^{(d-1)/2} \psi \right) - \frac{(d-1)(d-3)}{4r^2} \psi \right]
+  V(r)  \psi(r) \ =\ 0
  \ee
in $d$--dimensional space. We will assume that the spherically symmetric
potential $V(r)$ dies away at infinity as a power faster that $1/r^2$.
 Then, at large
$r$, our equation $\Delta \psi = 0$ has formally  two  solutions: {\it (i)}
$\psi(r) = $ const \ and  {\it (ii)} $\psi(r) \propto r^{2-d}$. The first
solution is not normalizable at infinity whereas the second one
(the Green's function of the laplacian in $d$ dimensions) {\it is} if
$d \geq 5$ (the measure  is $\int |\psi(r)|^2 r^{d-1} dr$).

Let us choose $d=5$. The solution of the free laplacian 
equation $\psi(r) \propto
r^{-3}$ is normalizable at infinity, but not at zero. Intuitively, it is
rather clear that, if the potential $V(r)$ is attractive and the well is
deep enough, a zero energy solution with the required asymptotics may be
found. Indeed, one can be easily convinced that the equation (\ref{Schr})
with the potential 
  \be
 \label{pot5}
V(r) \ =\ -\frac{15a^2}{2(r^2 + a^2)^2}
  \ee
has a nice normalized solution
  \be
  \label{psi5}
\psi(r) \ =\ \sqrt{\frac {2a}{\pi^3}} \ \frac 1 {(r^2 + a^2)^{3/2}}
 \ee
 Of course, the presence of the normalized at infinity zero energy solution
of the free Schr\"odinger equation  is just a {\it
necessary condition} for the existence
of  a zero energy
solution of the full Schr\"odinger equation normalized in the whole domain and
 does not guarantee it yet.
If the potential has other form than that in Eq.(\ref{pot5}), if it
is e.g. repulsive, Eq.(\ref{Schr}) has no solutions. If, however, this 
necessary condition is not satisfied, we can be sure that solutions are absent.
This would be  the case, for example, for the conventional 3--dimensional 
Schr\"odinger equation in the $s$--wave (The asymptotic normalizability
condition for the function $\propto r^{-l-1}$ is satisfied for 
$l \geq 1$ , and one can invent a 3--dim problem with the normalized
 zero energy solution in the $p$--wave, but it would not be a ground
state: the  $s$--wave states with negative energy would be present in 
 the spectrum.)

Let us return now to our SYM quantum mechanics. Only the theory with the
$SU(2)$ gauge group has been analyzed with this method so far, and we will
restrict ourselves to that case. The problem is supersymmetric and
the vacuum wave function satisfies not only the Schr\"odinger equation
$H|{\rm vac} \rangle = 0$, but also the equation 
$Q_\alpha|{\rm vac} \rangle = 0$. The necessary condition for a normalized
solution to this equation to exist is that the equation
  \be
  \label{Qslow0}
Q_\alpha^{\it eff} \chi_0^{\rm slow} (c_I, \mu_{\tilde \alpha}) \ =\ 0
  \ee
have a solution normalized at large $|{\bf c}|$. 

Following Ref. \cite{Hoppe1} but using the explicit form of the effective
supercharges in Eq.(\ref{QHeff}) which simplifies the reasoning and the 
derivation {\it a lot}, let us describe how such a solution for the ${\cal N} 
= 4$ theory can be constructed. The wave function
  \be
 \label{chicomp}
\chi_0^{\it slow} (c_I, \mu_{\tilde \alpha}) \ =\ 
a(c_I) + b_{\tilde \alpha} (c_I) \mu_{\tilde \alpha} + c_{\tilde \alpha 
\tilde \beta} (c_I) \mu_{\tilde \alpha} \mu_{\tilde \beta} + \ldots
  \ee
has altogether $2^8 = 256$ components. As was mentioned above, 
$\mu_{\tilde \alpha}$ do not provide a representation of $SO(9)$, but
the set of components $\{a(c_I), b_{\tilde \alpha} (c_I), \ldots \}$ does.
Indeed, if acting on the wave function (\ref{chicomp}) by the operator of spin
  \be
  \label{Spin}
S_{IJ} = \ \frac 14\ \lambda_\alpha 
(\Gamma_I \Gamma_J )_{\alpha \beta} \lambda_\beta
  \ee
 [with $\lambda_\alpha$ being expressed via $\mu_{\tilde \alpha}$
and $\bar \mu_{\tilde \alpha}$ according to Eq.(\ref{mulam})], we will
obtain again a function of the form (\ref{chicomp}). 

This representation
is reducible.
To understand it, note first that, when substituting (\ref{mulam}) in 
(\ref{Spin}), we obtain generically the terms of three types: $\propto
\bar \mu_{\tilde \alpha} \mu_{\tilde \beta}$, \ $\propto
 \mu_{\tilde \alpha} \mu_{\tilde \beta}$, and $\propto
\bar \mu_{\tilde \alpha} \bar \mu_{\tilde \beta}$. That means that, though
the fermion charge is not a good conserved quantity
\footnote{The precise meaning of this statement is the following.
{\it i)} The full hamiltonian (\ref{H}) does not commute with the full fermion
charge. {\it ii)} The effective hamiltonian in (\ref{QHeff}) commutes with the
``slow fermion charge'' $F^{\rm slow} =  \mu_{\tilde \alpha} \bar
\mu_{\tilde \alpha}$, but it does not help much 
 because we are in a position to 
solve Eq.(\ref{Qslow0}) involving the supercharge rather than hamiltonian, and the
commutator $[Q_\alpha, F^{\rm slow}]$ is a mess. The solution to 
 Eq.(\ref{Qslow0}) is not going to have a definite slow fermion charge.},
the {\it fermion parity} operator $(-1)^F$ is: it commutes with $S_{IJ}$ 
and anticommutes with the supercharge. Thus, a wave function involving only
even powers of $\mu_{\tilde \alpha}$ preserves its form under the action
of $S_{IJ}$. And so does the wave function involving only odd powers of
 $\mu_{\tilde \alpha}$. 

As it turns out, the latter presents an irreducible {\bf 128}--plet of
$SO(9)$. This is a kind of Rarita--Schwinger spin--vector 
$({\bf 128})_{I\alpha},\ I = 1,\ldots,9; \alpha = 1,\ldots,16$ satisfying 
the constraints $(\Gamma_I)_{\alpha\beta} ({\bf 128})_{I\beta} \ = 0$. The 
remaining
128 components of the wave function with $(-1)^F = 1$ split in two irreducible
representations 
$ {\bf 44} + {\bf 84}$. The first one
is the traceless symmetric tensor $({\bf 44})_{IJ}$ and the second one
is the antisymmetric tensor  $({\bf 84})_{IJK}$. 

Let us pick up the symmetric {\bf 44}--plet
\footnote{Its explicit form can be found in \cite{Hoppe1}.} and construct our 
asymptotic wave function as
  \be
  \label{psias}
\chi_0^{\rm slow} (c_I, \mu_{\tilde \alpha}) \propto
({\bf 44}^{\rm ferm})_{IJ}\  \partial_I \partial_J \frac 1{|{\bf c}|^7}
 \ee
It is an $SO(9)$ singlet. Obviously, it is normalizable at infinity. Acting
on it with the supercharge $Q_\alpha^{\rm eff}$, we obtain
  \be
  \label{actQchi}
Q_\alpha^{\rm eff} \chi_0^{\rm slow} (c_I, \mu_{\tilde \alpha}) 
\propto \ (\Gamma_K)_{\alpha\beta}
\lambda_\beta\  ({\bf 44}^{\rm ferm})_{IJ}\ 
 \partial_I \partial_J \partial_K
\frac  1{|{\bf c}|^7}
 \ee
The fermion structure in Eq.(\ref{actQchi}) is odd in $\mu_{\tilde \alpha}$
and presents our Rarita--Schwinger {\bf 128}--plet. We may write
  \be
 \label{Rarita}
(\Gamma_{\{K})_{\alpha\beta}
\lambda_\beta ({\bf 44}^{\rm ferm})_{IJ\}} \ =\ 
\delta_{IK} ({\bf 128}^{\rm ferm})_{J\alpha} +
\delta_{JK} ({\bf 128}^{\rm ferm})_{I\alpha} +
\delta_{IJ} ({\bf 128}^{\rm ferm})_{K\alpha} 
 \ee
Substituting it in Eq.(\ref{actQchi}), we obtain zero due to the property 
$\Delta (1/|{\bf c}|^7) = 0$. 

The existence of the normalized at infinity
solution to the equation (\ref{Qslow0}) is specific for the ${\cal N} = 4$
theory. Let us show that  no such solution exists in the 
 ${\cal N} = 1$ and  ${\cal N} = 2$ cases. Let first  ${\cal N} = 1$.
The situation is much simpler than for ${\cal N} =4$  because the fermion 
variables $\lambda_\alpha$ are complex  spinors in the representation 
{\bf 2} of the
$SO(3)$ group. They {\it are} the holomorphic variables on which the wave
function depends, and one need not bother to construct some other variables
like $\mu_{\tilde \alpha}$. The operator of the fermion charge
 $F = \lambda_\alpha \bar \lambda^\alpha$ commutes with the hamiltonian here
and the  properties  
\be
\label{comQF}
[Q_\alpha^{\rm eff}, F] =
- Q_\alpha^{\rm eff}, \ \ \ \
 [\bar Q^{\alpha \ {\rm eff}} , F] =
\bar Q^{\alpha \ {\rm eff}} 
  \ee
 hold.
 $2^2 =4$ components of the
wave function $\chi_0^{\rm slow} (c_i, \lambda_\alpha)$ are decomposed into
two singlets with $F = 0,2$ and a doublet with $F=1$. We see that a 
construction like in Eq.(\ref{actQchi}) is impossible and all the  solutions
to the equation (\ref{Qslow0}) have the form 
$\chi_0^{\rm slow} \ = \ {\rm const}(c_i)
f(\lambda_\alpha)$ and are not normalizable.

In the ${\cal N} = 2$ case, the fermion variables $\lambda_\alpha,\ \alpha =
1,\ldots,4$ belong to the representation {\bf 4} of $SO(5)$. Again, they 
are complex and can be chosen as holomorphic variables on which wave functions
depend. Again, $[H^{\rm eff}, F] = 0$, the properties (\ref{comQF}) are 
fulfilled, and one can look
for the solutions of Eq.(\ref{Qslow0}) in a sector with a particular $F$.
A 16--component wave function is decomposed into 2 singlets with $F = 0$ and
$F = 4$, 2 quartets with $F = 1,3$, and {\bf 5}--plet and a singlet with
$F = 2$. We can in principle construct the function
 \be
\label{N2try}
{\cal N} = 2: 
\chi_0^{\rm slow} (c_i, \lambda_\alpha) \ \stackrel {?}{=} \ 
\partial_j \frac 1 {|{\bf c}|^3} \ \lambda_\alpha (C 
\gamma_j)^{\alpha \beta}
\lambda_\beta \ ,
  \ee
where $\gamma_j$ are 5--dimensional 
gamma--matrices and $C$  is the 
charge conjugation matrix which lowers and rises spinor indices [when recalling
that $SO(5) \equiv Sp(4)$, $C$ is the antisymmetric skew--diagonal matrix
(\ref{Csympl})  
defining the group $Sp(4)$]. When acting 
on it with the effective supercharge $\bar Q^{\alpha\ {\rm eff}} =
\bar \lambda^\beta (\gamma_k)_\beta^{\ \alpha} \ \partial/\partial c_k$, 
we obtain
 $$
\partial_j \partial_k  \frac 1 {|{\bf c}|^3} \ 
\bar \lambda^\beta (\gamma_k)_\beta^{\ \alpha} \ \lambda_\gamma 
(C \gamma_j)^{\gamma\delta} \lambda_\delta \ =\ 
-2\partial_j \partial_k  \frac 1 {|{\bf c}|^3}\ 
\lambda_\gamma (C\gamma_j\gamma_k)^{\gamma\alpha}\ 
\propto \Delta\  \frac 1 {|{\bf c}|^3} \ =\ 0 \ .
 $$
The result of the action of the supercharge $Q_\alpha^{\rm eff}$ is also
zero. 

However, our best try (\ref{N2try}) {\it is} not an admissible solution 
because it does not satisfy the requirement of Weyl invariance 
(\ref{Weylsym}). 
Throwing it away, we are left with nothing.
Speaking of the ${\cal N} = 4$ effective wave function (\ref{psias}), it is 
Weyl invariant, is annihilated by effective supercharges and normalizable at 
infinity. It {\it is} the asymptotic solution we were looking for.
 
Obviously, the function (\ref{psias}) satisfies also the equation
   \be
  \label{Hslow0}
H^{\it eff} \chi_0^{\rm slow} (c_I, \mu_{\tilde \alpha}) \ =\ 0
  \ee
In the  ${\cal N} = 4$ theory, there is a unique Lorentz--invariant
function satisfying the equation (\ref{Hslow0}): this is how the solution
(\ref{psias}) was originally found \cite{Halp1}. Note, however, that
for  ${\cal N} = 2$ the equation  (\ref{Hslow0}) involves three
extra Weyl-- and Lorentz--invariant solutions:
$\chi_0^{\rm slow} = 1/|{\bf c}|^3$, 
$\chi_0^{\rm slow} = \lambda_\alpha C^{\alpha \beta} \lambda_\beta /|{\bf c}|^3 $, and $\chi_0^{\rm slow} = 
(\lambda_\alpha C^{\alpha \beta} \lambda_\beta)^2 /|{\bf c}|^3 $. These 
effective wave functions are not annihilated by the effective supercharges
and do not correspond to any normalized supersymmetric vacuum state in
the full theory.

It would be interesting to generalize this analysis for other groups.
In particular, for the groups $Sp(2n \geq 6)$, $SO(n \geq 8)$ and for the
exceptional groups, the full theory
has several normalized solutions and hence the asymptotic equations 
(\ref{Qslow0}), (\ref{Hslow0}) should have
 several normalized  Weyl--invariant solutions. A good educated guess
is that, for  the {\it supercharge} equation,
the inverse is also true and {\it any} Lorentz-- and Weyl--invariant 
solution to the  equation (\ref{Qslow0}) can be promoted up
 to a normalized supersymmetric vacuum in the full theory.

\section{Conclusions.}
We have discussed three different techniques which allow one to deduce the 
existence of the normalizable supersymmetric vacuum states in the 
${\cal N} = 4$ SYM quantum mechanics: {\it (i)} the  method of mass 
deformation, {\it (ii)} the functional integral method, and {\it (iii)}  
the asymptotic wave function method. The mass deformation method is, of course,
the most straightforward and the simplest one. It allowed us to obtain the 
result and determine the number of normalized states for {\it all} gauge 
groups.

But two other methods are also interesting and valuable. First, it is really
thrilling to see how the {\it completely} different ways of reasoning give the 
identical
results for the physical Witten index whenever the comparison is possible
(at the moment, the existence of a supersymmetric vacuum state
for the $SU(2)$ group is observed with all three methods
and with the  methods {\it (i)}  and {\it (ii)} --- for higher
unitary groups.)
Second, the method {\it (i)} involves a possible weak point: we are sure of
existence of the quantum supersymmetric vacuum (vacua) for
large mass, but cannot {\it prove} that all the states we have found stay 
normalizable in the limit $M \to 0$. It might happen in principle that  one 
or several or all such states become delocalized at this point. We do not
find it probable, but independent confirmation of our result by other methods,
especially with the functional integral method which is the most bullet--proof
is highly desirable.
   
We are indebted to A. Elashvili, N. Nekrasov, and A. Vainshtein for 
illuminating discussions. 

{\bf Note added.} It appears that there is also the fourth method 
to calculate $\#_{\rm vac}$ which uses the D--brane
language and ideology. In the very recent \cite{Hanany}, our results for 
 $\#_{\rm vac}$ in the case of  symplectic and orthogonal gauge groups were 
reproduced in this way.

\section*{Appendix. Principal contribution for $SU(2)$.}
\renewcommand{\theequation}{A.\arabic{equation}}
\setcounter{equation}0

We will describe here how the results (\ref{princres2}) are obtained
and
explain the reasons of disagreement between the results of the old
\cite{jaIW} and new \cite{YiSethi} calculations.

The starting point of Ref.\cite{jaIW} was the Cecotti--Girardello
formula (\ref{CG}) for the index. However, this formula cannot be 
directly applied to the gauge theories where the degrees of freedom 
forming the physical phase space are not explicitly singled out. To
use Eq.(\ref{CG}), we have first to resolve the Gauss' law constraints.
 Consider the simplest ${\cal N} = 1$ theory with the hamiltonian
 \be 
 \label{HN1}
H = \frac 12 E_j^A E_j^A + \frac {g^2}4 
(\epsilon^{ABC}  A^B_j A^C_k )^2 +
ig \epsilon^{ABC} \lambda^A_\alpha (\sigma_j)^\alpha_{\ \beta} 
\bar\lambda^{\beta B} A^C_j \ ,
 \ee
where $j = 1,2,3$ and $\lambda^A_\alpha$,\  $\alpha = 1,2$, \ are holomorphic
fermion variables. The constraint is
  \be
 \label{Gauss1}
G^A = \epsilon^{ABC}\left(A_j^BE_j^C  -  i \lambda_\alpha^B  
\bar\lambda^{\alpha C}  \right) \ =\ 0\ .
  \ee
To separate the physical degrees of freedom, it is convenient to use the
polar representation
  \be
 \label{polarULV}
A^A_j \ =\ U_{jk} \Lambda^B_k V^{BA} \ ,
  \ee
where $U$ and $V$ are orthogonal matrices and $\Lambda^B_k = {\rm diag}
(a,b,c)$. We have 6 physical variables: $a,b,c$, and  3 Euler angles 
$\theta,\phi,\psi$ of the matrix $U$, and 3 gauge degrees of freedom ---
the Euler angles of the matrix $V$. The hamiltonian (\ref{HN1}) can be 
rewritten via the new variables as follows:
   \be
  \label{IW6}
H \ =\ \frac 12 \left\{
p_a^2 + p_b^2 + p_c^2 + \frac{(b^2 + c^2)(I_1^2 + J_1^2) + 4bcI_1 J_1}
{(b^2 - c^2)^2} \right.\nonumber \\ \left.
+  \frac{(a^2 + c^2)(I_1^2 + J_1^2) + 4acI_1 J_1}
{(a^2 - c^2)^2} \ +\  \frac{(a^2 + b^2)(I_3^2 + J_3^2) + 4abI_3 J_3}
{(b^2 - a^2)^2} \right \} \nonumber \\
+ \frac 12 g^2 (a^2b^2 + a^2c^2 + b^2c^2 ) +  
ig \epsilon^{ABC} \lambda^A_\alpha (\sigma_j)^\alpha_{\ \beta} 
\bar\lambda^{\beta B} A^C_j \ ,
  \ee
where $p_a, p_b, p_c$ are the canonical momenta of the variables $a,b,c$;
$I_i$ are the standard combinations representing the generators of $SO(3)$
which depend on the Euler angles $\theta,\phi,\psi$ 
and their canonical momenta :
   \be 
  \label{IW7}
 I_1 \ &=&\ \sin \psi\ p_\theta + \cos \psi \left( \cot \theta \ p_\psi
- \frac{p_\phi}{\sin \theta}  \right)\ , \nonumber \\
I_2 \ &=&\ \cos \psi\ p_\theta - \sin \psi \left( \cot \theta \ p_\psi
- \frac{p_\phi}{\sin \theta}  \right)\ , \nonumber \\
  I_3 \ &=&\ p_\psi\ ,
  \ee
and $J^A$ are the analogous combinations for the matrix $V$.

The major advantage of the decomposition (\ref{polarULV}) is a great 
simplification of the constraints. When expressed via new variables,
they acquire the following nice form
  \be
 \label{IW8}
J^A \ =\ i\epsilon^{ABC} 
\lambda^{B}_{ \alpha} \bar\lambda^{ \alpha C}\ .
  \ee
Substituting it in Eq.(\ref{IW6}), we obtain the gauge--fixed classical
hamiltonian depending only on physical variables. We can go now with this
hamiltonian to Eq.(\ref{CG}) where the integration is performed over the
physical phase space
  \be
 \label{IW9}
 \prod_n \frac {dx_n dp_n}{2\pi}
\prod_a d\bar\psi_a d\psi_a \ =\ \frac 1{(2\pi)^6} da dp_a db dp_b dc dp_c
dU dI_1 dI_2 dI_3  \prod_{A\alpha} d\bar\lambda^{\alpha A} d\lambda_\alpha^A\ ,
  \ee
where $dU = sin \theta d\theta d\phi d\psi$. Defining carefully the proper
range of integration over the variables $a,b,c,\theta, \phi,\psi$, 
inserting into the integral the unity
  \be
  \label{IW10}
1 \ =\  \int \prod_A dJ^A \delta (J^A - i\epsilon^{ABC} 
\lambda^{B}_{ \alpha} \bar\lambda^{\alpha C}) \nonumber \\
=\ \left( \frac {\beta g }{2\pi} \right)^3 
\prod_A \int dJ^A dA_0^A  \ \exp\left\{i\beta g A_0^A (J^A - i\epsilon^{ABC} 
\lambda^{B}_{ \alpha} \bar\lambda^{\alpha C}) \right\}
  \ee
(so that the non--dynamic variables $A_0^A$ entering the original
Yang--Mills lagrangian are restored), and calculating the fermion determinant,
the integral can be expressed eventually in the symmetric form 
(\ref{indprinc}) (with ${\cal N} =1$), but with the extra overall factor
4. Calculating the integral gives $I_W = 1$ which is 4 times larger
than the correct value $I_W = 1/4$.

The resolution of this paradox is the following.  $I_W = 1$ {\it is}
the correct value for the Cecotti--Girardello integral (\ref{CG}) for the
index of the gauge--fixed hamiltonian $H^{\rm g.f.}$ obtained when 
substituting Eq.(\ref{IW8}) into Eq.(\ref{IW6}). However, $H^{\rm g.f.}$
is not {\it completely} equivalent to the original hamiltonian (\ref{HN1})
with the constraints (\ref{Gauss1}). 
Note that $H^{\rm g.f.}$ enjoys the ${\bf Z}_2 \times {\bf Z}_2$ discrete
symmetry 
\footnote{Using the low case letters for the indices of 
$\lambda_\alpha^{a,b,c}$
reflects the fact that they are treated now as fermion superpartners
to the bosonic variables $a,b,c$ rather than as components of a colored
vector. But, of course, $\lambda_\alpha^a \equiv \lambda_\alpha^A$.}
  \be
 \label{Z2Z2}
(a,b,c,\ \lambda_\alpha^a, \lambda_\alpha^b, \lambda_\alpha^c) \ 
&\longrightarrow& \ 
(-a,-b,c,\ -\lambda_\alpha^a, -\lambda_\alpha^b,\lambda_\alpha^c) \nonumber \\
  (a,b,c,\ \lambda_\alpha^a, \lambda_\alpha^b, \lambda_\alpha^c) \ 
&\longrightarrow& \ 
(-a,b,-c,\ -\lambda_\alpha^a, \lambda_\alpha^b,-\lambda_\alpha^c) \nonumber \\
(a,b,c,\ \lambda_\alpha^a, \lambda_\alpha^b, \lambda_\alpha^c) \ 
&\longrightarrow& \ 
(a,-b,-c,\ \lambda_\alpha^a, -\lambda_\alpha^b,-\lambda_\alpha^c) \ .
   \ee
The symmetry (\ref{Z2Z2}) presents a remnant of the original gauge symmetry
[it corresponds to multiplying $A_j^A,\ \lambda_\alpha^A$ by the orthogonal
matrices $V^{AB} \ =\ {\rm diag} (-1,-1,1),\ {\rm diag} (-1,1,-1)$ and
diag$(1,-1,-1)$ ] and plays exactly the same role as the Weyl symmetry
of the effective hamiltonian in Eq.(\ref{QHeff}) discussed at length in Sect. 4
\footnote{It that case the  symmetry was not  ${\bf Z}_2 \times {\bf Z}_2$
, but just  ${\bf Z}_2 $ because we were interested with abelian classical
vacuum configurations for which two of three elements of the diagonal matrix
$\Lambda_k^B$ vanish.}. The states non-invariant under the symmetry
(\ref{Z2Z2}) are not physical ones and should be discarded. 

We know already from Sect. 4 how to implement such a discrete symmetry for
the path  integral for the index. In the full analogy with Eqs.(\ref{IndK2}, 
\ref{IndK}), we may write
   \be
 \label{IndK4}
 I_W \ =\ \frac 1{4} \int da db dc dU 
\prod_{{ \alpha} } d \bar \lambda^{a \alpha} d\lambda_{ \alpha}^a 
 d \bar \lambda^{b \alpha} d\lambda_{ \alpha}^b 
 d \bar \lambda^{c \alpha} d\lambda_{ \alpha}^c 
\exp
\left\{ -   \bar \lambda^{a \alpha} \lambda_{\alpha}^a -
\bar \lambda^{b \alpha} \lambda_{ \alpha}^b - 
\bar \lambda^{c \alpha} \lambda_{ \alpha}^c  \right  \} \nonumber \\
\left[
{\cal K} ( a,b,c,  \bar  \lambda^{a \alpha },  \bar \lambda^{b \alpha }, 
\bar \lambda^{c \alpha }; \ a,b,c,    \lambda^a_{\alpha },  
\lambda^b_{\alpha }, 
 \lambda^c_{\alpha } ;\ \beta) \ +  \right. \nonumber \\ \left.
{\cal K} ( -a,-b,c,  -\bar\lambda^{a\alpha},  -\bar\lambda^{b\alpha}, 
\bar \lambda^{c\alpha}; \ a,b,c,    \lambda_{\alpha}^a,  \lambda_{\alpha}^b, 
 \lambda_{\alpha}^c ;\ \beta)
+ \ {\rm two\ more \ terms}  \right] \ . 
 \ee
Consider the second integral. The fermion part of the evolution operator \\
${\cal K} (\ldots,  -\bar\lambda^{a\alpha},  -\bar\lambda^{b\alpha}, 
\bar \lambda^{c\alpha}; \ \ldots ,    \lambda_{\alpha}^a,  \lambda_{\alpha}^b, 
 \lambda_{\alpha}^c;\ \beta )$ involves the factor
$\exp \{ -   \bar \lambda^{a \alpha} \lambda_{\alpha}^a -
\bar \lambda^{b \alpha} \lambda_{ \alpha}^b + 
\bar \lambda^{c \alpha} \lambda_{ \alpha}^c    \}$ [cf. Eq.(\ref{kernel})] so
that we obtain the overall factor 
$\propto \exp \{ -   2\bar \lambda^{a \alpha} \lambda_{\alpha}^a -
2\bar \lambda^{b \alpha} \lambda_{ \alpha}^b    \}$ in the measure. As it 
was the case for the calculation of the first term in Eq.(\ref{IndK4}), it
is convenient to insert the unity as in Eq.(\ref{IW8}) so that the Lorentz
invariance is restored. The fermion determinant is now
  \be
 \label{detAB}
\det \| i\beta g (\sigma_\mu)^\alpha_\beta \epsilon^{ABC} A^C_\mu -
2\delta_\beta^\alpha {\rm diag}(1,1,0) \| \ \sim\ 
(\beta g)^4 [(A_\mu^1)^2 + (A_\mu^2)^2 ]^2 \ .
  \ee
The bosonic part of the evolution operator involves now the extra factor
$\propto \exp\{ -2(a^2 + b^2) /\beta \}$. Performing the same transformations
as in \cite{jaIW}, we arrive at the integral
 \be 
\label{interm}
\sim \frac{(\beta g)^3 (\beta g)^4}{\beta^{9/2}} \int \prod_\mu
dA^1_\mu dA^2_\mu dA^3_\mu \ [(A_\mu^1)^2 + (A_\mu^2)^2 ]^2 \nonumber \\
\exp \left\{ - \frac 2\beta [(A_\mu^1)^2 + (A_\mu^2)^2 ] \right \} 
\exp \left\{ - \frac {\beta g^2}2 (A^3_\mu)^2 [(A_\mu^1)^2 + (A_\mu^2)^2 ]^2
\right \} \ .
  \ee
The last factor comes from the potential where we have neglected 
the small terms $\sim (A^{1,2}_\mu)^4$ . This integral is estimated as
$\propto \beta^{9/2} g^3 $ which vanishes together with two other terms
in Eq.(\ref{IndK4}) in the limit $\beta \to 0$.
Thus, the final result for the index in the ${\cal N} = 1$ theory is 
   \be
\label{arrow}
I_W = 1 \ \stackrel{{\bf Z}_2 \times {\bf Z}_2}{\longrightarrow} \ 
I_W = \frac 14 [1+0+0+0] \ =\ \frac 14\ .
  \ee

We see that the requirement of invariance of the wave functions with respect 
to the symmetry (\ref{Z2Z2}) has reduced the value for the index fourfold.
This is related to the fact that the principal contribution alone does not 
count the number of normalized vacuum states, but is contaminated by the 
contribution of the continuum spectrum states. 

In is instructive to consider a simple example of the system with discrete
spectrum where this phenomenon {\it does} not happen. Take the oscillator
with the hamiltonian (\ref{Hosc}). The hamiltonian and the supercharges
are invariant under the transformation $(x \to -x, \psi \to -\psi)$. The
index of the system involving only the states even under this symmetry
is given by the integral
  \be 
\label{intosceven}
I_W \ =\ \frac 12 \int  dx  d\bar\psi d\psi e^{-\bar\psi \psi}
\left[
{\cal K} ( x,  \bar  \psi  ;\ x, \psi;\ \beta) \ +
{\cal K} ( -x,  -\bar  \psi  ;\ x, \psi;\ \beta)
 \right] \nonumber \\
 =\  \frac 12 + \frac 12 \int \frac{dxdp}{2\pi}
d\bar\psi d\psi e^{-2\bar\psi \psi}
\exp\left\{- \frac{\beta p^2}2 + 2ipx \right\} \ =\ \frac 12 + \frac 12 \ =\ 
1\ ,
 \ee
the same as with Eq.(\ref{intind}).  That could not be otherwise, of course,
because the value $I_W = 1$ corresponds to the presence of a supersymmetric
bosonic vacuum state invariant under the transformation $x,\psi \to -x,-\psi$.

But for the system in interest, the terms coming from the integration of 
non--diagonal ${\cal K}(\cdots)$ vanish and the result is given by 
Eq.(\ref{arrow}).
 
This result was obtained in \cite{YiSethi} by another method. Instead of 
resolving
explicitly the Gauss law constraints and then coming to grips with 
implementing carefully the residual discrete gauge symmetry, we could write
  \be
\label{gavr}
I_W \ =\ \frac 1{8\pi^2} \int dV \int \prod_{Aj} dA^A_j \prod_{A \alpha}
d\bar \lambda^{A\alpha} d\lambda_\alpha^A \
{\cal K} (A^B_j V^{BA}, \bar \lambda^{B\alpha} V^{BA}; \ 
A^A_j ,  \lambda^{A}_\alpha ;\ \beta)\ ,
 \ee
where ${\cal K}(\cdots)$ is now the evolution operator of the 
{\it unconstrained}
hamiltonian (\ref{HN1}) (and $8\pi^2$ is the volume of the $SO(3)$ gauge 
group). The integral (\ref{gavr}) automatically takes into account only
the gauge--invariant states in the spectral decomposition of  
${\cal K}(\cdots)$. It is reduced to the integral (\ref{indprinc}) with
a correct prefactor. 

For the ${\cal N} = 2$ and  ${\cal N} = 4$ theories, the analogs of 
Eq.(\ref{gavr}) lead to the result (\ref{indprinc}) with the correct
 coefficient. And the same result can be obtained by resolving explicitly
the Gauss law constraints and implementing the discrete symmetry
${\bf Z}_2 \times {\bf Z}_2$.
We have repeated the calculations of Ref.[2b] and obtained the results
$I_W[H^{\rm g.f.}] = 1$ for the ${\cal N} = 2$ theory and 
$I_W[H^{\rm g.f.}] = 5$ for the ${\cal N} = 4$ theory which differ from the
original results of Ref. [2b] by a factor of 2. Dividing it further
by $\#({\bf Z}_2 \times {\bf Z}_2) \ =\ 4$, we reproduce the result
 (\ref{indprinc}).

\end{document}